\documentclass[twocolumn]{aastex631}

\shorttitle{}
\shortauthors{}

\graphicspath{{./}{figures/}}

\usepackage{placeins, amsmath, mathtools, physics, booktabs, threeparttable, bm, natbib, hyperref, float}
\begin{document}

\title{Do We Have Sufficient Knowledge of the Galactic Foreground Emission in Cosmic Microwave Background Science?}

\correspondingauthor{Hao Liu, Yi-Fu Cai}
\email{ustc\_liuhao@163.com, yifucai@ustc.edu.cn}


\author[0000-0003-3172-3714]{Jia-Rui Li}
\affiliation{Department of Astronomy, School of Physical Sciences, University of Science and Technology of China, No.96, JinZhai Road, Baohe District, Hefei, Anhui 230026, China.}
\affiliation{CAS Key Laboratory for Researches in Galaxies and Cosmology, School of Astronomy and Space Science, University of Science and Technology of China, Hefei, Anhui, 230026, China.}
\affiliation{School of Physics and optoelectronics engineering, Anhui University, 111 Jiulong Road, Hefei, Anhui, China 230601.}
\affiliation{Beijing Institute of Tracking and Telecommunication Technology, Beijing 100094, China.}

\author[0009-0009-5702-8788]{Peibo Yuan}
\affiliation{Department of Astronomy, School of Physical Sciences, University of Science and Technology of China, No.96, JinZhai Road, Baohe District, Hefei, Anhui 230026, China.}
\affiliation{CAS Key Laboratory for Researches in Galaxies and Cosmology, School of Astronomy and Space Science, University of Science and Technology of China, Hefei, Anhui, 230026, China.}
\affiliation{School of Physics and optoelectronics engineering, Anhui University, 111 Jiulong Road, Hefei, Anhui, China 230601.}

\author[0000-0003-0706-8465]{Yi-Fu Cai}
\affiliation{Department of Astronomy, School of Physical Sciences, University of Science and Technology of China, No.96, JinZhai Road, Baohe District, Hefei, Anhui 230026, China.}
\affiliation{CAS Key Laboratory for Researches in Galaxies and Cosmology, School of Astronomy and Space Science, University of Science and Technology of China, Hefei, Anhui, 230026, China.}

\author[0000-0003-4410-5827]{Hao Liu}
\affiliation{School of Physics and optoelectronics engineering, Anhui University, 111 Jiulong Road, Hefei, Anhui, China 230601.}

\begin{abstract}
Galactic foreground emission plays a key role in cosmic microwave background (CMB) science, particularly for detecting primordial gravitational waves.
A well-known lesson is the ``dust wave'' identified by BICEP2 in 2014, which was ruled out through a more careful analysis of foreground emission.
To date, most estimates of Galactic foreground emission have relied on the assumption that for each line of sight, only one component is considered per emission mechanism. However, the results in this work suggest that more complex modeling---particularly involving multiple components arising from either line-of-sight complexity or pixel mixing---may be necessary to fully account for Galactic foregrounds, including dust and other components. 
More interestingly, the only available two-component dust estimate also fails due to oversimplified emission parameters, although it is conceptually superior to single-component alternatives.
These results yield three key conclusions:
(1) Due to the intrinsic three-dimensional complexity of the Galactic environment, where physical conditions vary with both distance and direction, the actual radiation from Galactic foreground components cannot be accurately characterized by single-component models. 
(2) Consequently, CMB experiments require more frequency bands to resolve these components. 
(3) Spatial variations of foreground emission parameters should \emph{not} be simplified, because in this work, all such simplifications are found to degrade the estimates significantly.
\end{abstract}



\section{Introduction} 
\label{sec:intro}

The investigation of the cosmic microwave background (CMB) has been integral to the evolution of modern cosmology, playing a crucial role in milestones such as the validation of the Big Bang theory \citep{1965ApJ...142..414D}, the progress in understanding primordial nucleosynthesis \citep{1973ApJ...179..343W}, and the refinement of the $\rm{\Lambda}$CDM model \citep{1996ApJ...464L...1B, 2003ApJS..148..175S, 2020A&A...641A...6P, 2020PhRvD.101h3504I}. 
Notably, the primordial CMB $B$-mode polarization, which is exclusively produced by primordial gravitational waves, offers a distinctive avenue for testing the validity of inflationary cosmology \citep{1997PhRvL..78.2054S}.

Unfortunately, foreground contamination presents challenges in the critical frequency range (70---220 GHz) for CMB studies.
For frequencies below 80 GHz, synchrotron and free-free emissions must be taken into account, while thermal dust emission becomes increasingly dominant above 80 GHz \citep{2013A&A...553A..96D, 2020A&A...641A...1P}.
Additionally, anomalous microwave emission (AME) affects frequencies around 70 GHz~\citep{2014A&A...565A.103P}, and the cosmic infrared background (CIB) introduces further interference at frequencies above 300 GHz~\citep{2016A&A...596A.109P}.
In the context of polarization, free-free emission is generally regarded as negligible~\citep{2014PTEP.2014fB109I, 2024A&A...691A.110L}, whereas the polarization of thermal dust emission can exhibit significant complexity~\citep{2024arXiv241112801C}. 

Due to the presence of atmospheric water vapor, the atmospheric windows available to ground-based CMB telescopes are discontinuous across the microwave frequency range \citep{2015ApJ...809...63E}. 
Thus, space-based observatories offer a significant advantage in estimating foreground emission parameters, particularly those related to thermal dust emission. 
Following the RELIKT-1 project \citep{1992MNRAS.258...71K}, which initiated space-based observations of the CMB, the COBE satellite pioneered measurements of CMB anisotropies \citep{1992ApJ...397..420B}, paving the way for subsequent missions such as WMAP and \textit{Planck} to advance precision cosmology.
WMAP operates across five frequency bands from 23 to 94 GHz, but lacks high-frequency coverage \citep{2003ApJ...583....1B}. 
In contrast, \textit{Planck} extends the frequency range up to 857 GHz with its six High-Frequency Instrument (HFI) bands (100---857 GHz, \citealp{2020A&A...641A...3P}) and three Low-Frequency Instrument (LFI) bands (30, 44, and 70 GHz, \citealp{2020A&A...641A...2P}), enabling more precise measurements of thermal dust and CIB emissions~\citep{2016A&A...596A.109P}.
LiteBIRD, a next-generation mission, will further expand observations across 15 frequency bands (40---402 GHz) using three telescope sets: LFT, MFT, and HFT (\citealp{2018JLTP..193.1048S, 2018SPIE10698E..1YS}). 
Its key channels (100, 119, and 140 GHz) are designed to overlap with each other, reducing systematic errors and enhancing constraints on additional dust components such as AME~\citep{2023PTEP.2023d2F01L}.
PICO, another proposed space mission, will feature 21 frequency bands spanning 21---799 GHz and is designed to deliver superior sensitivity and angular resolution \citep{2019arXiv190210541H, 2019LPICo2135.5035T}. 
With its extensive frequency coverage, PICO can enable more detailed modeling of foreground components and their polarization properties.

However, the modeling of Galactic foregrounds in CMB research still faces several challenges:
(1) Multiple foreground components including synchrotron, free-free emission, and AME, exhibit similar spectral behaviors at low frequencies~\citep{2016A&A...594A..10P}, leading to degeneracies that hinder an accurate component separation. 
(2) Current models of AME heavily rely on empirical fitting, with limited grounding in well-established physical mechanisms~\citep{2016A&A...594A..10P}, leaving considerable uncertainties. 
(3) The polarization of synchrotron and dust emission is spatially complex and poorly constrained~\citep{2020A&A...641A..11P}, resulting in significant modeling uncertainties that challenge the extraction of faint CMB primordial $B$ modes~\citep{2017MNRAS.468.4408H}. 
Furthermore, the issue of multi-component thermal dust and synchrotron emission remains critical in polarization. Unlike temperature mode where intensities add up monotonically, the polarization signal is highly sensitive to the superposition of multi-component if they have misaligned polarization angles. 

Recently, we introduced a novel approach to address the complexity of foreground emission along the line of sight \citep{2025ApJS..276...45L}. 
This approach investigates the relationship between linear regression ratios (noted as $R$) and cross-correlation coefficients (noted as $C$) for sky patches observed at different frequencies.
We pointed out that although the expectations of $R$ and $C$ are independent, the deviation of $C$ from 1 and the error of $R$ are tightly linked. 
This connection enables a purely data-driven test of the model-to-data mismatch, which has ruled out existing single-component estimates of Galactic thermal dust emissions.
For convenience, a brief methodology explanation can be found in Appendix~\ref{app:method}.

In this paper, we extend our analysis to all accessible foreground estimates in the CMB channels, including both two-component thermal dust emission models and available single-component estimates of other emission mechanisms, such as synchrotron, AME, and free-free emission.
The goal is to determine whether any of them is consistent with the observational data, or at least approaches the acceptance threshold.

The structure of this work is straightforward: 
We briefly review the methodology in Appendix~\ref{app:method} and present results in Sections~\ref{sec:thermal dust emission}---\ref{sec: test of other components} (covering thermal dust, synchrotron, and other emission mechanisms). 
Conclusions are given in Section~\ref{sec:conclusions}, with reproducible software available at GitHub,\footnote{\url{https://github.com/liuhao-cn/foreground_validation}} which can also be used to evaluate consistency between any foreground model and observational data.

\section{Test of thermal dust emission}
\label{sec:thermal dust emission}
Since thermal dust emission dominates above 150 GHz, \textit{Planck}'s primary data products for this emission mechanism are sky maps at 353, 545, and 857 GHz. 
They exhibit significantly higher signal-to-noise ratios for thermal dust than other components/bands, making thermal dust the most studied foreground emission mechanism to date.
Consequently, we initiate our analysis by evaluating all thermal dust emission models that we can find. 
Compared to \citet{2025ApJS..276...45L} that tested only single-component modified blackbody models, this section expands the evaluation to include all major thermal dust models in current use, particularly two-component and physical modeling approaches, as listed in Table~\ref{tab:models} and explained below:
\begin{enumerate}
\item Analytical models:
This class contains single- and double-component modified blackbody models.
\item Physical models:
This class is based on the electromagnetic radiation properties of various dust grains.
\item 3D model:
This class incorporates the 3D distribution of dust clouds in analysis.
\end{enumerate}
\begin{table}[!hbt]
\centering
\caption{List of thermal dust estimates to be tested.}
\begin{threeparttable}
\begin{tabular*}{0.43\textwidth}{cccc} 
\toprule
~~~~~~~~~~&Type &&~~~~~~~~~~~model~~~~~~~~~~~ \\
\midrule
&Class I &~~~& Planck13\tnote{a} \\
 &&~~~& Planck15-C\tnote{b} \\
 &&~~~& Planck15-G\tnote{c} \\
 &&~~~& SRoll\tnote{d} \\
 &&~~~& Irfan19\tnote{e} \\
 &&~~~& Meisner\tnote{f} \\
\hline
&Class II &~~~& DL07\tnote{g}\\
 &&~~~& HD17-d5\tnote{h,i}\\
 &&~~~& HD17-d7\tnote{h,i} \\
\hline
&Class III &~~~& 3D\tnote{j} \\
\bottomrule
\end{tabular*}
\begin{tablenotes}
\footnotesize
\item[a] \cite{2014A&A...571A..11P}
\item[b] \cite{2016A&A...594A..10P}
\item[c] \cite{2016A&A...596A.109P}
\item[d] \cite{2021A&A...650A..82D}
\item[e] \cite{2019A&A...623A..21I}
\item[f] \cite{1999ApJ...524..867F, 2015ApJ...798...88M}
\item[g] \cite{2007ApJ...663..866D, 2007ApJ...657..810D}
\item[h] \cite{2017ApJ...836..179H, 2017ApJ...834..134H}
\item[i] \url{https://pysm3.readthedocs.io/en/latest/models.html}
\item[j] \cite{2018MNRAS.476.1310M}
\end{tablenotes}
\end{threeparttable}
\label{tab:models}
\end{table}

\subsection{Class I}
\label{sec:class-I}
\begin{figure*}[htpb]
\centering
\includegraphics[width=\linewidth]{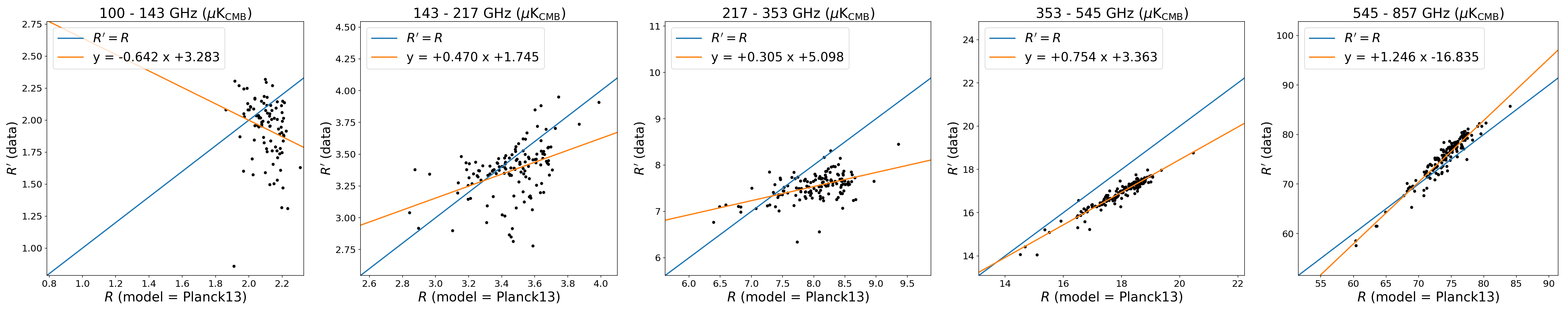}
\includegraphics[width=\linewidth]{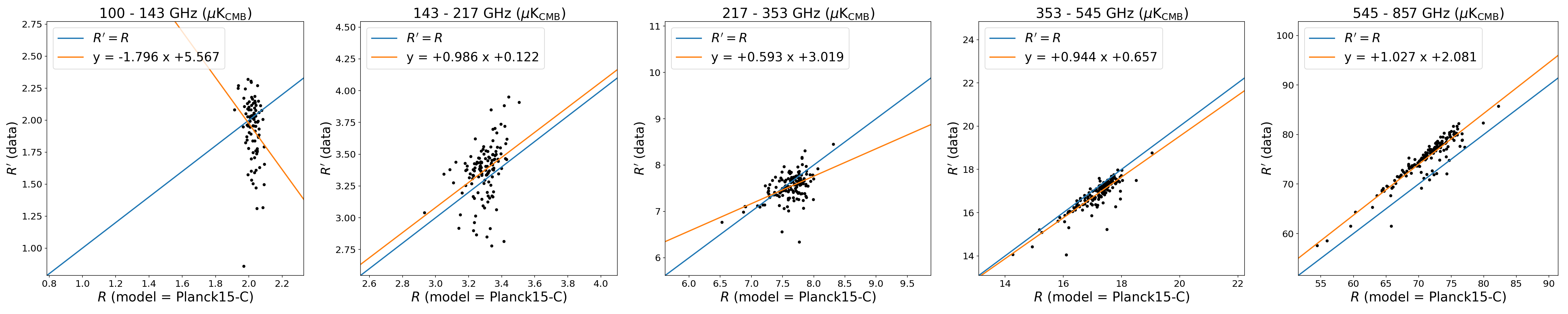}
\includegraphics[width=\linewidth]{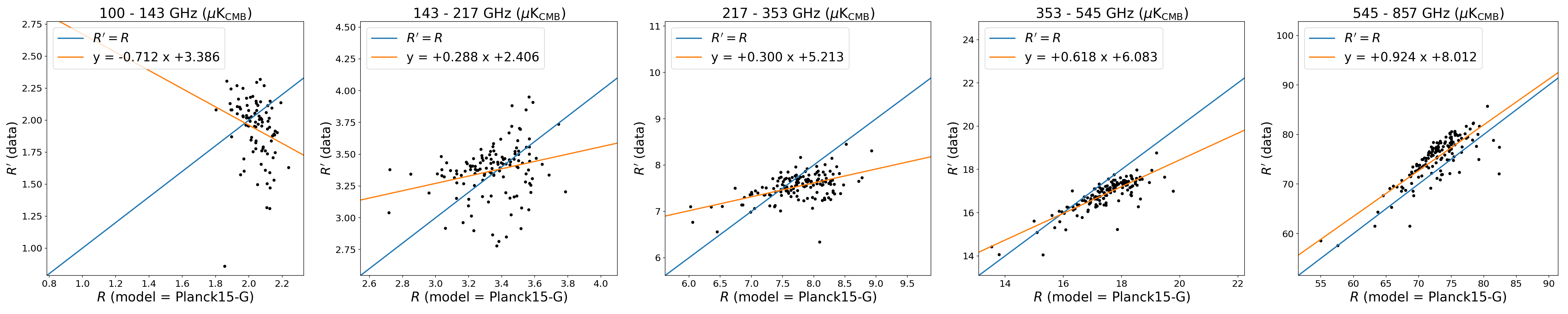}
\includegraphics[width=\linewidth]{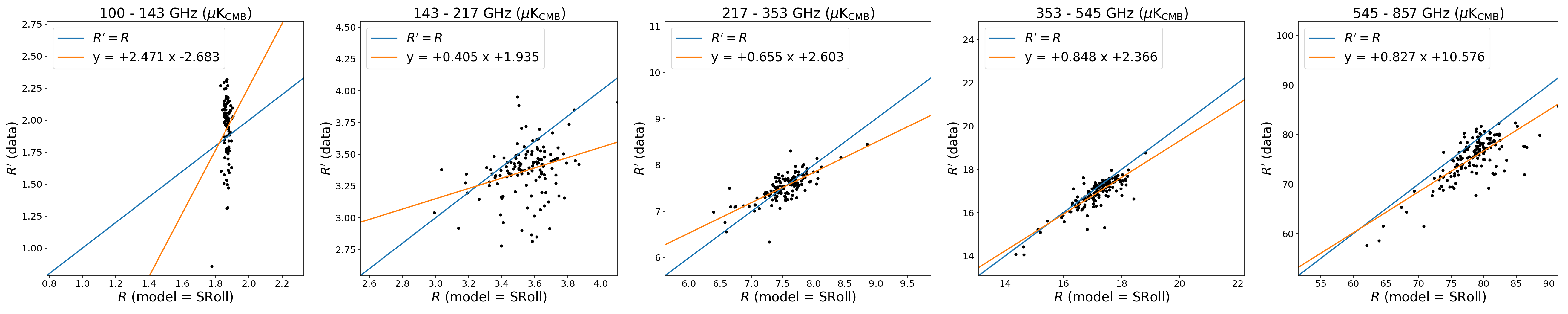}
\includegraphics[width=\linewidth]{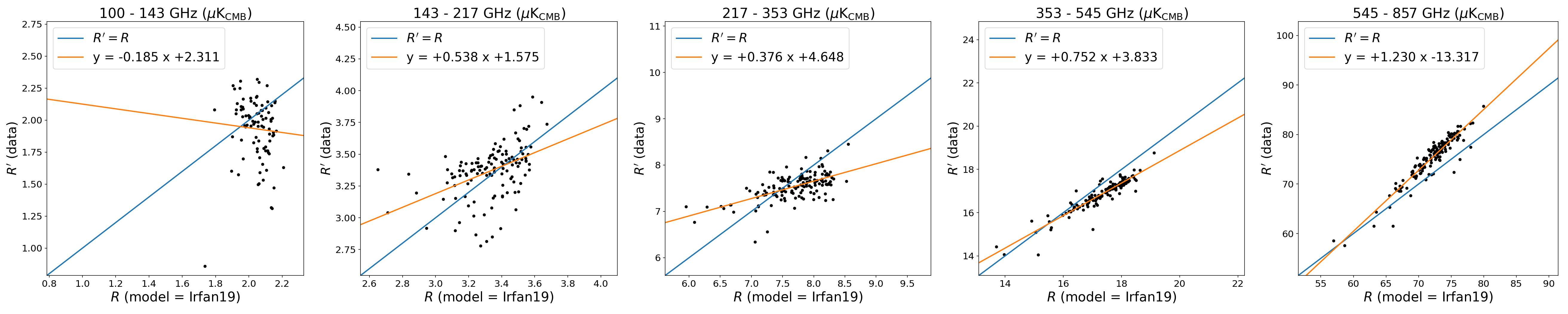}
\includegraphics[width=\linewidth]{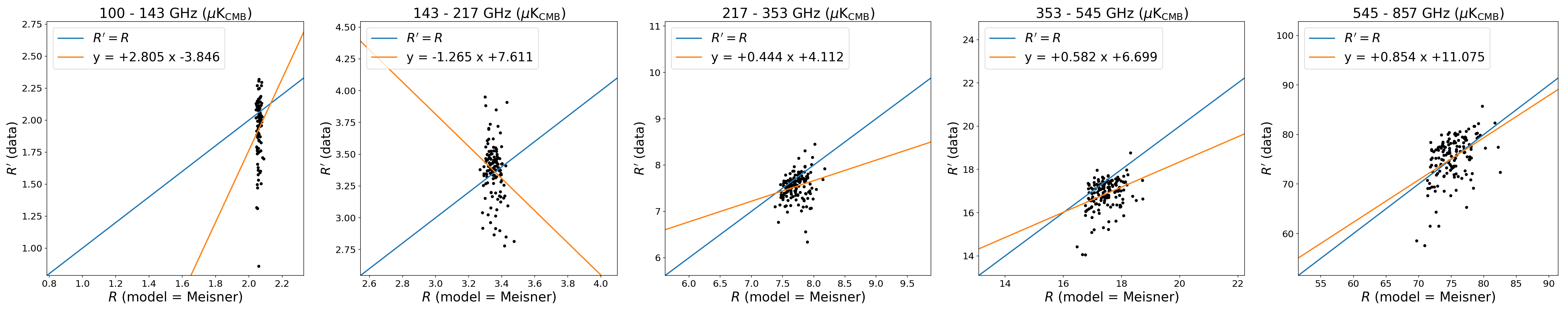}
\caption{
Scatter plots comparing linear regression ratios $R^\prime$ (vertical, from data) with $R$ (horizontal, from model) for Section~\ref{sec:class-I}. 
Rows 1---6: Model Planck13, Planck15-C, Planck15-G, SRoll, Irfan19, and Meisner.
The blue lines are the expectations with $R^\prime = R$ while the orange lines represent the least squares regression of the scatter points.
}
\label{fig:scatter-class-I}
\end{figure*}

For optically thin interstellar media, including interstellar dust, the radiative transfer equation takes the form: 
\begin{equation}
I(\nu) = I_{0}(\nu) \Big[1 - \tau_\mathrm{d}\left(\nu\right)\Big] + \tau_\mathrm{d}\left(\nu\right) B_\nu(T),
\end{equation}
where $B_\nu(T)$ represents the Planck's function, 
$I(\nu)$ and $I_{0}(\nu)$ signify the observed and primordial emission intensity, typically expressed in $\mathrm{MJy\cdot sr^{-1}}$. 
The first term on the right-hand side represents the transmitted background radiation, and the second term corresponds to thermal dust emission, $I_\mathrm{d}(\nu)$. 
At long-wavelength-limit, the dust optical depth $\tau_\mathrm{d}\left(\nu\right)$ can be approximated by a power-law function of frequencies \citep{2011piim.book.....D}; thus, thermal dust emission can be modeled as: 
\begin{equation}
I_\mathrm{d}(\nu) = \tau_\mathrm{d}(\nu_0) \left(\frac{\nu}{\nu_0}\right)^{\beta_\mathrm{d}} B_\nu(T), 
\end{equation}
which is known as the (single-component) ``modified blackbody spectrum model.'' 
When multiple dust components with different physical properties are considered, the emission can be generalized as: 
\begin{equation}\label{equ:td main}
I_\mathrm{d}(\nu) = \sum_{i=1}^N \tau_\mathrm{d}^i(\nu_0) \left(\dfrac{\nu}{\nu_0}\right)^{\beta^i_\mathrm{d}} B_\nu(T^i_\mathrm{d}),
\end{equation}
where $\tau_\mathrm{d}^i(\nu_0)$, $T_\mathrm{d}^i$, and $\beta_\mathrm{d}^i$ correspond to the $i$-th component's reference optical depth, thermal equilibrium temperature, and spectral index, respectively. 

As discussed in Section~\ref{sec:intro}, the limited number of frequency bands available in high-frequency microwave sky surveys leads to the widely adopted single-component assumption, which requires only three degrees of freedom per line of sight.
We will briefly introduce the context of each model in this category, followed by a unified assessment and comparative analysis. 

\paragraph{\textbf{Model \textit{Planck}13}}
Basing on the \textit{Planck} 2013 release, \cite{2014A&A...571A..11P} adopts a single-component dust emission model (hereafter referred to as Planck13) using $\chi^2$ minimization at $5'$ resolution from \textit{Planck}'s 2013 CMB-free maps at 353, 545, and 857 GHz~\citep{2014A&A...571A...1P} along with the IRAS 3000 GHz map \citep{1984ApJ...278L...1N}. 
In the context of Equation~(\ref{equ:td main}), this corresponds to $N=1$ and their reference frequency for thermal dust emission is set to  $\nu_0 = 353\,\mathrm{GHz}$. 

\paragraph{\textbf{Model \textit{Planck}15-C}}
Along with the \textit{Planck} 2015 release, \citet{2016A&A...594A..10P} introduces an other single-component modified blackbody model (hereafter referred to as Planck15-C), with a resolution of $60'$, utilizing the Bayesian Commander analysis framework \citep{2004ApJS..155..227E, 2006ApJ...641..665E, 2008ApJ...676...10E}.
This model incorporates the full suite of \textit{Planck} frequency channels, and the reference frequency for thermal dust is $\nu_0 = 545\,\mathrm{GHz}$:
\begin{equation}
I_\mathrm{d}(\nu) = I_\mathrm{d}(\nu_0) \left(\dfrac{\nu}{\nu_0}\right)^{\beta_\mathrm{d}} \dfrac{B_\nu(T)}{B_{\nu_0}(T)}.
\end{equation}
Unlike Planck13 that only focuses on thermal dust emission, Planck15-C incorporates all major foreground components and utilizes all available frequency bands for estimation.
This comprehensive approach enables a useful comparison to Planck13.

\paragraph{\textbf{Model \textit{Planck}15-G}}
\label{subsection:Planck15_G}
Both Planck13 and Planck15-C are influenced by the CIB emission \citep{2016A&A...594A..10P}.
To address this issue, \textit{Planck} team employs the Generalized Needlet Internal Linear Combination (GNILC) method to reduce the CIB contamination from the 353, 545, and 857 GHz CMB-free maps.
This process produces improved dust emission maps with a resolution of $5'$, which are subsequently used to derive a new suite of single-component dust emission parameters (\citealp{2016A&A...596A.109P}, hereafter referred to as Planck15-G).

\paragraph{\textbf{Model SRoll}} 
Recently, \cite{2021A&A...650A..82D} calibrates the \textit{Planck} maps using precise measurements of the large-scale, full-frequency-range differential solar system kinetic dipole.
By employing a momentum-expanded, single-component modified blackbody spectrum as follows: 
\begin{equation}
I(\nu) = I(\nu_0)f(\nu,\beta) + \dfrac{\partial f}{\partial\beta}\mathrm{\Delta}I_{1}(\nu_0) + \dfrac{\partial^2 f}{\partial\beta^2}\mathrm{\Delta}I_{2}(\nu_0), 
\label{equ:SRoll}
\end{equation}
where $f(\nu,\beta)$ is the modified blackbody spectrum, 
they generate large-scale thermal dust emission maps smoothed at $60'$ (referred to as the SRoll estimate hereafter) across the \textit{Planck} frequency bands ranging from 100 to 857 GHz. 
These maps inherently include color corrections, significantly enhancing their usability. 
However, SRoll assumes a constant global dust temperature of $T = 18\,\mathrm{K}$, which inevitably limits the model's ability to describe dust spectrum's variation across different directions.

\paragraph{\textbf{Model Irfan19}}
In addition to the aforementioned models, \cite{2019A&A...623A..21I} utilizes a sparsity-based parametric approach to develop an alternative single-component dust model at a $5'$ resolution (referred to as model Irfan19 hereafter).

\paragraph{\textbf{Model Meisner}}
\label{sec:two-component}
This is the only two-component estimate of thermal dust emission to date. 
The authors claimed that modeling thermal dust emission in the microwave band using a single-component approach cannot fully capture the underlying physical conditions \citep{2015ApJ...798...88M}.
This limitation likely arises from the intricate physical and chemical composition of interstellar dust, which comprises amorphous graphite, silicates, polycyclic aromatic hydrocarbons (PAHs), and even iron-bearing compounds \citep{1965ApJ...142.1681S, 1985MNRAS.215..425R, 2002ApJS..138...75P, 2011piim.book.....D, 2019arXiv190210541H}.
Furthermore, dust grains of varying compositions exhibit distinct size distributions \citep{1977ApJ...217..425M, 1984ApJ...285...89D, 2019ApJ...887..244T} and electromagnetic emission characteristics \citep{1984ApJ...285...89D, 2001ApJ...554..778L, 2007ApJ...657..810D}.
Among these components, carbon-based and silicon-based grains are most prevalent and significant.

With the aforementioned understanding, \cite{1999ApJ...524..867F} introduced a two-component dust model, which was subsequently refined by \citet{2015ApJ...798...88M}, referred to as model Meisner hereafter. 
They subtracted CMB anisotropies, other foreground signals, and dipole components from the \textit{Planck} 2013 release \citep{2014A&A...571A...6P} and the DIRBE/IRAS map \citep{1998ApJ...500..525S}.
Subsequently, they fitted the two-component modified blackbody model to the remaining signal, which is likely dominated by thermal dust emissions: 
\begin{equation}
I_\mathrm{d}(\nu) \propto f_1 q_1 \left(\frac{\nu}{\nu_0}\right)^{\beta^1_\mathrm{d}} B_\nu(T^1_\mathrm{d}) + (1-f_1) q_2 \left(\frac{\nu}{\nu_0}\right)^{\beta^2_\mathrm{d}} B_\nu(T^2_\mathrm{d}), 
\label{equ:Meisner}
\end{equation}
where $q_i$ denotes the ratio of the far-infrared emission cross section to the effective absorption cross section for the $i$-th species \citep{1999ApJ...524..867F}, and $f_1$  represents the fractional contribution of the first component to the total dust emission intensity.
Following the approach of \cite{1999ApJ...524..867F}, thermal equilibrium between the two components is assumed to simplify the estimation.
In addition, the limited number of frequency bands forces the use of global constants for certain parameters, including $f_1 = 0.0485$, $q_1/q_2 = 8.219$, $\beta_1 = 1.63$, and $\beta_2 = 2.82$. 

Figure~\ref{fig:scatter-class-I} demonstrates the performance of aforementioned modified blackbody spectrum dust emission models across the frequency range of 100---857 GHz, by evaluating the consistency of the ratios $R$ and $R'$ derived from models and data, respectively.
If the model and data are in agreement, the scatter points should align closely with the blue line ($R=R'$), thereby providing a clear visual indication of the model-to-data discrepancies. 

At frequencies above 353 GHz, single-component modified blackbody models, i.e., Planck13, Planck15-C, Planck15-G, SRoll, and Irfan19, demonstrate reasonably good performance. 
However, significant model-to-data deviations are observed in the 100---353 GHz range. 
In particular, the 100---143 GHz pair exhibits a scatter plot where the orange regression line deviates substantially from the expectation. 
Additionally, the horizontal distribution of dust ratio $R$ (from model) is notably narrower compared to the vertical distribution of $R^\prime$ (from data). 
Consequently, all these models are deemed inaccurate and cannot be reliably extrapolated from high frequencies to infer the amplitude of dust emission at intermediate frequencies (100---143 GHz), even approximately. 

As illustrated in rows 1---3 of Figure~\ref{fig:scatter-class-I}, the range of $R$ for Planck15-C is considerably narrower compared to Planck13 and Planck15-G, suggesting that, despite the inclusion of more bands and components in the data processing, the estimation of the dust spectrum's variation is actually less accurate in Planck15-C. 
This suggests that incorporating additional foreground components without robust experimental support, such as an adequate number of frequency bands and good signal-to-noise ratio, could deteriorate the results. 
Additionally, the results are marginally inferior for Planck15-G in the 143---545 GHz range, which could potentially be attributed to inaccuracies in CIB removal or subtle discrepancies between the GNILC and Commander pipelines.
As for SRoll, the limitation in spatial variation is precisely captured by our method, as demonstrated in row 4 of Figure~\ref{fig:scatter-class-I} for the 100---143 GHz band, where a significant suppression of $R$ along the horizontal axis is clearly observed.
Such an error can hardly be mitigated through higher-order expansions of other parameters.

The performance of model Meisner is particularly interesting. 
In principle, this model incorporates two dust components, which is definitely much better than single component. 
However, its performance is notably inferior to any single-component model, as demonstrated in row 6 of Figure~\ref{fig:scatter-class-I}.
A key indicator of this issue is the near absence of spatial variation in the ratio $R$, as evidenced by the extremely narrow distribution of points along the horizontal direction within the 100---217 GHz range. 
A similar problem is also evident in the SRoll model, as illustrated in row 4 of Figure~\ref{fig:scatter-class-I}, for the same reason. 
These findings underscore the critical importance of rigorously testing any simplifications in dust emission estimation to prevent unintended consequences.
In particular, it is \emph{strongly advised} to prohibit the use of a constant dust temperature or spectral index, as they can lead to significant side effects.

Other issues are also observed in the 217---857 GHz range for model Meisner, where the alignment between the regression lines and the scatter plots is significantly poorer compared to that of single-component models.
This further confirms the fact that using constant parameters introduces more problems than benefits.

Although \cite{2016A&A...594A..10P} notes that they cannot effectively distinguish between single- and two-component models without higher-frequency observations, our approach demonstrates clear distinctions among these dust models, even within the existing frequency coverage. 
This reveals the sensitivity of our method and highlights its potential to enhance the testing of dust models. 
It is worth noting that there remains some debate in the literature regarding whether the spectral energy distribution of thermal dust emission exhibits spatial variations. 
For instance, \citet{2017A&A...599A..51P} reported spatial variations in the polarized spectral energy distribution of thermal dust emission, suggesting that such de-correlation could arise from either spatial variations in the thermal dust spectrum or from changes in the dust polarization angle. 
\citet{2018PhRvD..97d3522S} revisited the data processing approach of \citet{2017A&A...599A..51P}, introducing several improvements in the analysis, and concluded that there is no statistically compelling evidence for de-correlation of thermal dust emission in the \textit{Planck} release. 
\citet{2020A&A...641A..11P}, using a model-dependent MCMC fitting method, didn't detect frequency de-correlation in thermal dust emission between 100 and 353 GHz, 
though the study noted that its results do not rule out the possibility of such de-correlation.  
Methodologically, our approach diverges from the aforementioned studies by operating entirely in the pixel domain, as opposed to the power spectrum domain. 
Some other studies such as \citet{2019PTEP.2019c3E01I} have presupposed spatial variations in the foreground spectral energy distribution and focused on developing mitigation techniques based on this assumption. 
Our investigation addresses a more foundational question: whether a single-component model suffices to characterize these spatial variations in the first place. 

\begin{figure}[htpb]
\centering
\includegraphics[width=0.4\textwidth]{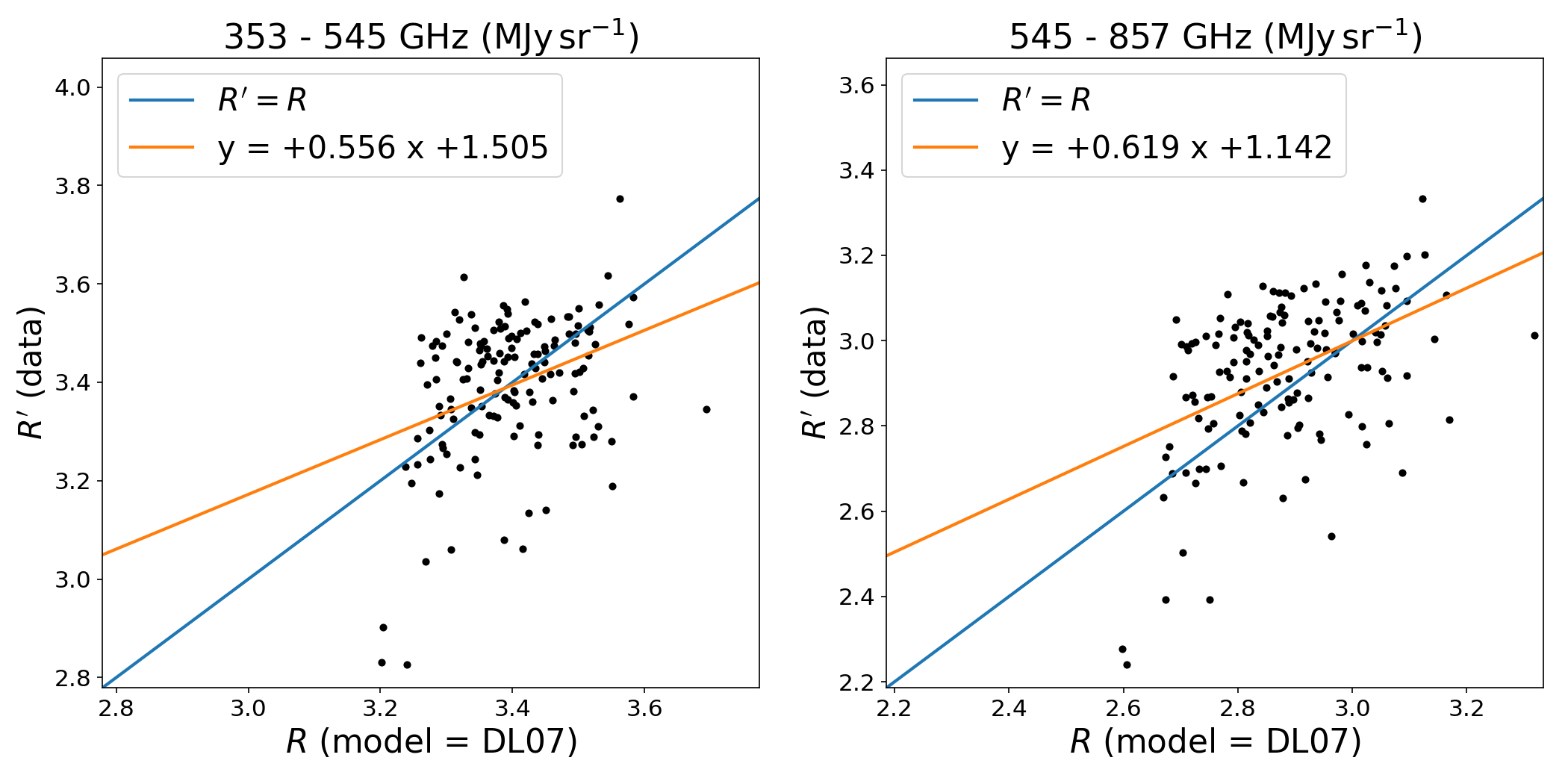}
\caption{Similar to Figure~\ref{fig:scatter-class-I} but for model DL07 with only 353---857 GHz (due to data availability). }
\label{fig:scatter_DL07}
\end{figure}

\subsection{Class II and III}
\label{sec:physical models}

\begin{figure*}[htpb]
\centering
\includegraphics[width=\textwidth]{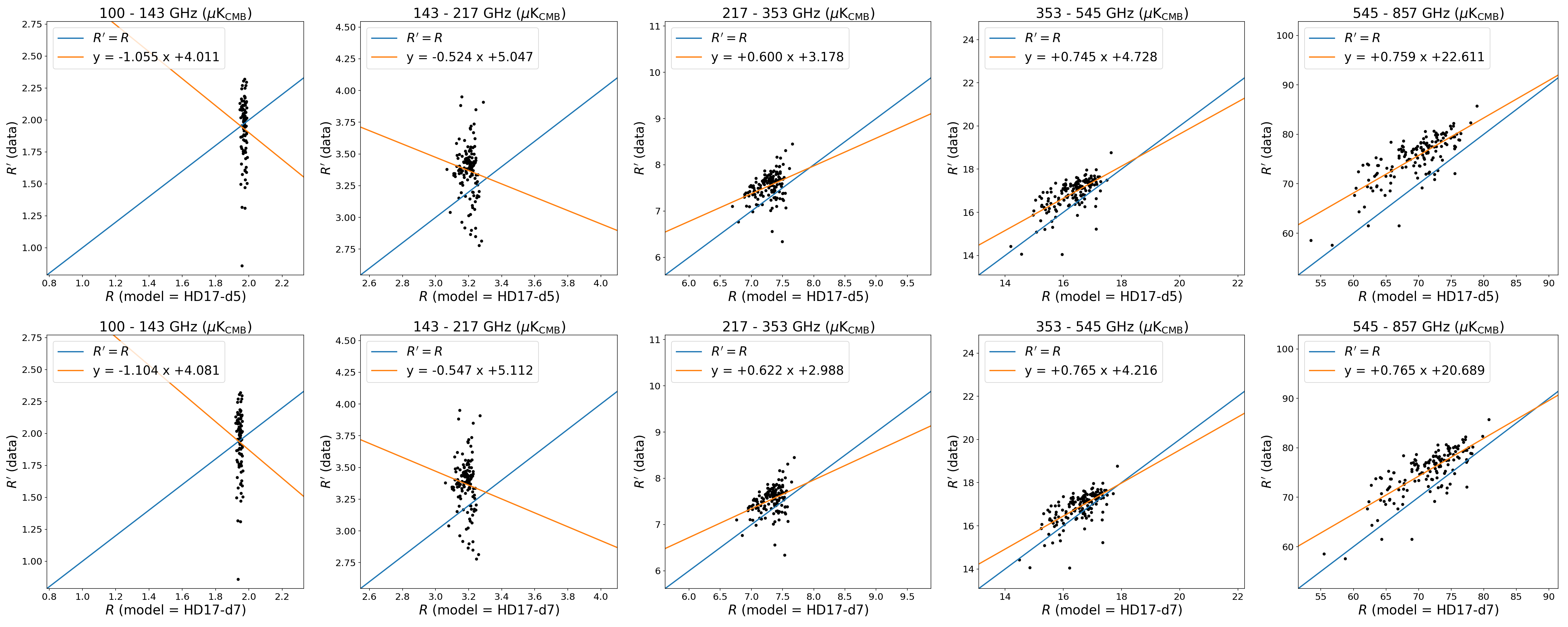}
\caption{Same as Figure~\ref{fig:scatter-class-I} but for HD17-d5 (upper) and HD17-d7 (lower). }
\label{fig:scatter_HD17}
\end{figure*}

\begin{figure*}[htpb]
\centering
\includegraphics[width=\textwidth]{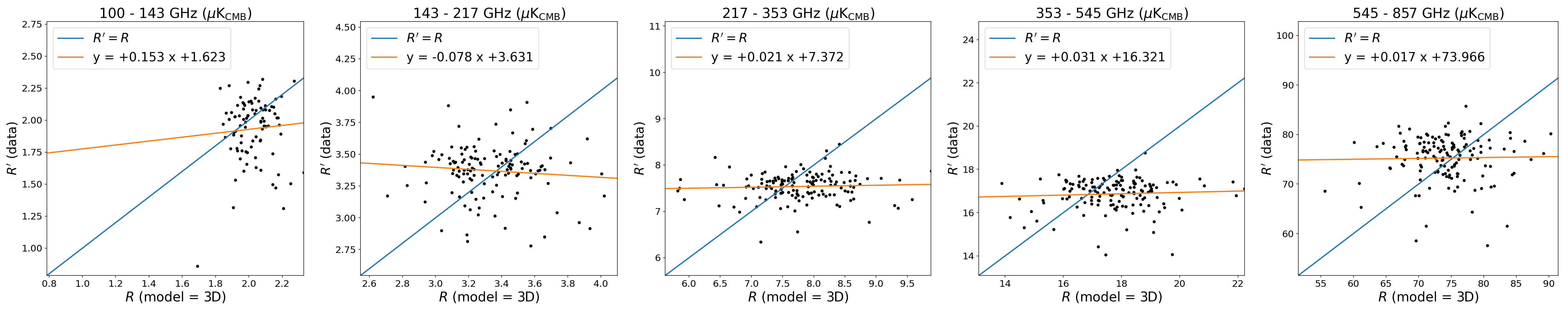}
\includegraphics[width=\textwidth]{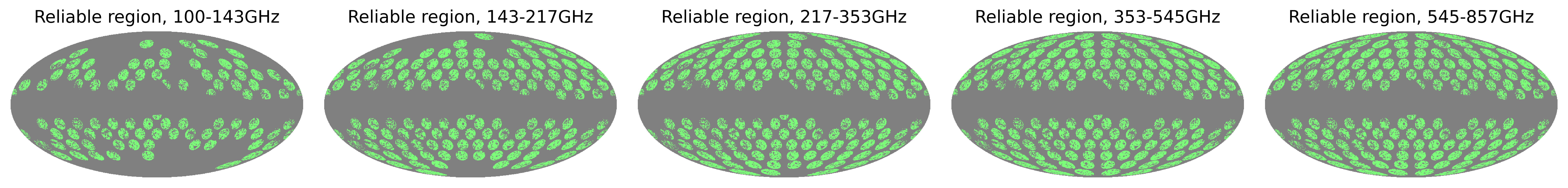}
\caption{
Top panel: same as Figure~\ref{fig:scatter-class-I} but for the 3D model. 
Bottom panel: reliable regions with $C^\prime\geq0.95$ in green. 
These regions exhibit universal applicability for validating all thermal dust models, as their delineation is fundamentally observation-driven. 
}
\label{fig:scatter_3D}
\end{figure*}

This subsection continue evaluating the performance of the class II (physical models) and class III (3D model). 
We begin with class II, which are based on the physical/chemical properties of interstellar dust. 

As already mentioned, it is widely recognized that Galactic dust predominantly consists of carbonaceous and silicate grains, with a possible minor contribution from PAHs and trace amounts of iron-containing particles \citep{1965ApJ...142.1681S, 1985MNRAS.215..425R, 2002ApJS..138...75P, 2019arXiv190210541H}.
The emission properties of dust grains, which are influenced by their chemical composition and physical size/shape, can be partially deduced through astronomical observations and laboratory measurements \citep{1984ApJ...285...89D, 2001ApJS..134..263W, 2001ApJ...551..807D, 2001ApJ...554..778L, 2002ApJ...576..762L, 2012ApJ...757..103D}.
This fundamental understanding forms the basis for the development of physical models of thermal dust emission, which are intrinsically linked to the properties of the dust grains.
However, it is evident that the corresponding measurement and calculation are highly challenging.

\paragraph{\textbf{Model DL07}}
This is a silicate-graphite-PAHs model developed by \cite{2007ApJ...663..866D} and \cite{2007ApJ...657..810D} to elucidate the composition and emission properties of dust grains.
This model constrains the PAH abundance in dust ($q_\mathrm{PAH}$) and the parameters of the interstellar radiation field in which the dust is embedded ($f_\mathrm{PDR}$ and $U_{\min}$).
\cite{2016A&A...586A.132P} applied the DL07 model to the \textit{Planck} 2015 data release \citep{2016A&A...594A...1P}, as well as the IRAS \citep{1984ApJ...278L...1N} and WISE \citep{2010AJ....140.1868W} observations, further refining the model parameters.

Figure~\ref{fig:scatter_DL07} compares the maps of DL07 estimate at \textit{Planck} frequencies of 353, 545, and 857 GHz with the observed dust data maps,
\footnote{\href{https://irsa.ipac.caltech.edu/data/Planck/release_2/all-sky-maps/previews/COM_CompMap_Dust-DL07-ModelFluxes_2048_R2.00}{COM\_CompMap\_Dust-DL07-ModelFluxes\_2048\_R2.00.fits}}
with all units standardized to $\mathrm{MJy\cdot sr^{-1}}$. 
As illustrated in this figure, the agreement with expectations is poor, even at these high-frequency bands.

\paragraph{\textbf{Models HD17-d5 and HD17-d7}}
In contrast to the dust composition described in the DL07 model, \cite{2017ApJ...836..179H} and \cite{2017ApJ...834..134H} explore the radiative properties of dust models that exclude PAHs.
The \texttt{PySM3} software package \footnote{\url{https://pysm3.readthedocs.io/en/latest/}} \citep{2017MNRAS.469.2821T, 2021JOSS....6.3783Z, 2025arXiv250220452T} has implemented this model with two parametrization schemes: HD17-d5 (excluding iron) and HD17-d7 (including iron-bearing components).
In this study, we employ \texttt{PySM3} to compare them at \textit{Planck} frequencies in Figure~\ref{fig:scatter_HD17}, revealing negligible differences between these two models.

In summary, Figures~\ref{fig:scatter_DL07} and \ref{fig:scatter_HD17} demonstrate that existing physical models, which rely on the electromagnetic emission properties of dust grains, exhibit significant limitations, even within the dust-dominated frequency range of 353---857 GHz.
However, this does not imply that the foundational theoretical framework is flawed.
Instead, it highlights the intricate complexity of interstellar dust, indicating that our observation, understanding and the reductionist approaches employed remain insufficient to precisely characterize dust emission at this stage.

\paragraph{\textbf{Model 3D}}
To more accurately capture the three-dimensional nature of dust distribution, \cite{2018MNRAS.476.1310M} proposes a three-dimensional dust emission model incorporating polarization.
This model accounts for spatial variations in dust density, spectral index, and thermal equilibrium temperature along each line of sight.
Galactic dust is divided into six layers, with each layer represented as a single-component modified blackbody.
Thus, the dust emission's Stokes parameters $I$, $Q$, and $U$ share the same spectral index and thermal equilibrium temperature within a given layer.
To reproduce small-scale ($<1^\circ$) dust polarization signals, this model introduces random perturbations on small scales.
However, it is important to note that, due to significant limitations in the number of available observational frequency bands, the authors do \emph{not} assert the model to be exact.

Figure~\ref{fig:scatter_3D} demonstrates obviously inferior results for the 3D model, which is not surprising.
These findings reveals the limitations in dataset rather than model itself.
Consequently, the corresponding results should be considered preliminary and interpreted with caution.

In summary, the limitations of current dust models mainly arise from the inherent complexity of thermal dust emission.
This complexity is driven not only by the diverse chemical composition of dust grains, which influences their emission spectra and polarization properties; 
but also by the intricate three-dimensional spatial distribution of dust clouds.
For dust clouds located at different distances along one line of sight, local thermal equilibrium can probably be achieved, but a complete equilibrium across all distances is almost impossible.
In principle, dust parameters must vary with distance, yet accurately determining these variations remains a big challenge. 
For example, \cite{2018MNRAS.476.1310M} provides a three-dimensional distribution of dust emission but does not include temperature-specific details.
Moreover, the moment expansion method discussed by \citet{2025A&A...697A.212V} demonstrates that this three-dimensional complexity will present substantial challenges for future CMB observations. 
In addition to these challenges, extragalactic foregrounds further complicate the precision of CMB measurements \citep{2024arXiv241115307H, 2024arXiv241024026T}. 

Finally, regarding the number of thermal dust components and three-dimensional distribution of dust components, our current analysis constrains medium-to-large angular scales. 
Crucially, observational signatures represent integrated line-of-sight superpositions, while the referenced ``scale'' inherently denotes angular resolution on the celestial sphere. 
Consequently, our results demonstrate negligible dependence on the specific selection of three-dimensional physical scale. 
However, comprehensive modeling incorporating the full three-dimensional dust distribution---particularly small-scale perturbations---could reveal component numbers substantially exceeding current estimates.
Critically, Earth's location within the Milky Way is non-special. 
Therefore, variations in dust parameters measured along great-circle trajectories (angular paths) provide statistically valid estimators for variations along physical sight lines. 
From the perspective of three-dimensional spatial statistics, these measurement paths are fundamentally indistinguishable. 
Parameter variations derived from angular trajectories thus offer observationally tractable proxies for line-of-sight fluctuations.
Moreover, since conventional dust parameters represent line-of-sight averages, this methodology systematically underestimates the true amplitude of parameter variations along individual sight lines.

\section{Test of synchrotron emission}
\label{sec: test of sync}

\begin{figure*}[htpb]
\centering
\includegraphics[width=0.8\textwidth]{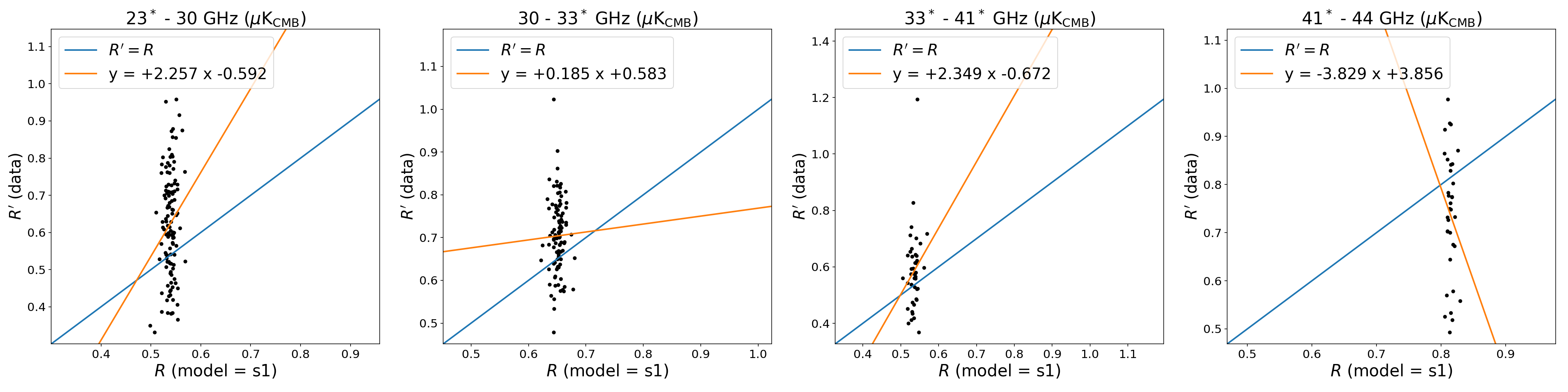}
\includegraphics[width=0.8\textwidth]{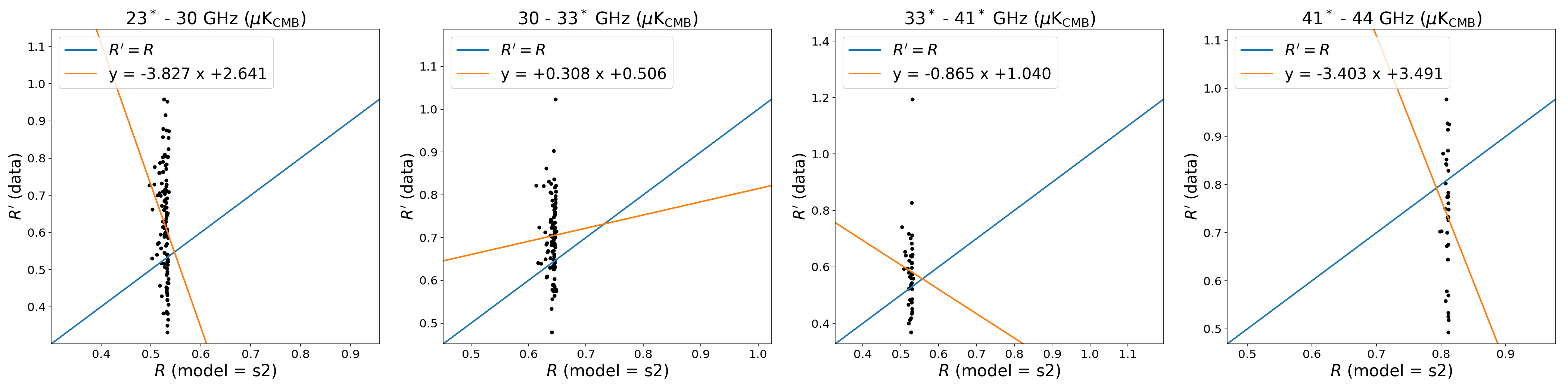}
\includegraphics[width=0.8\textwidth]{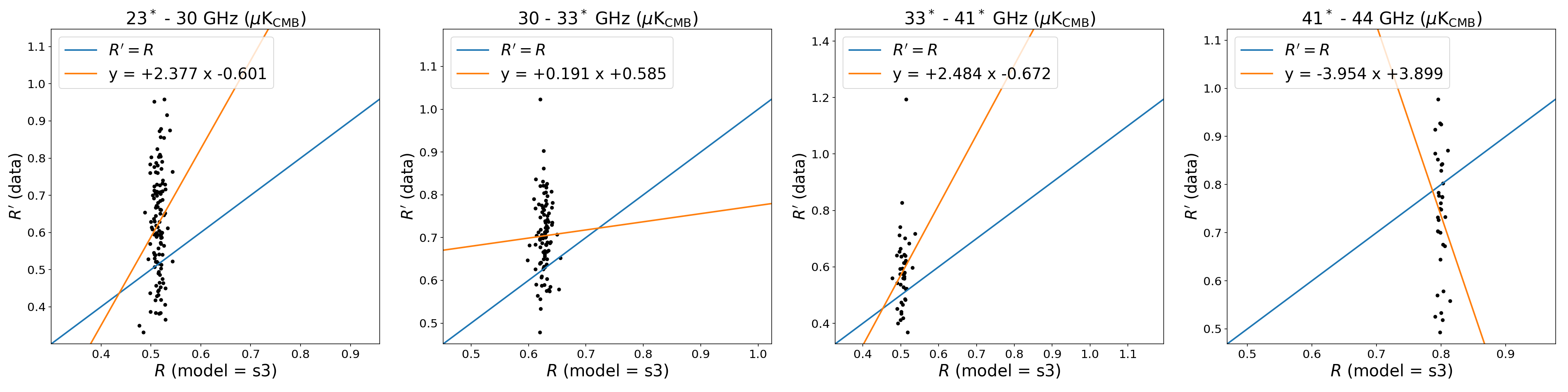}
\includegraphics[width=0.8\textwidth]{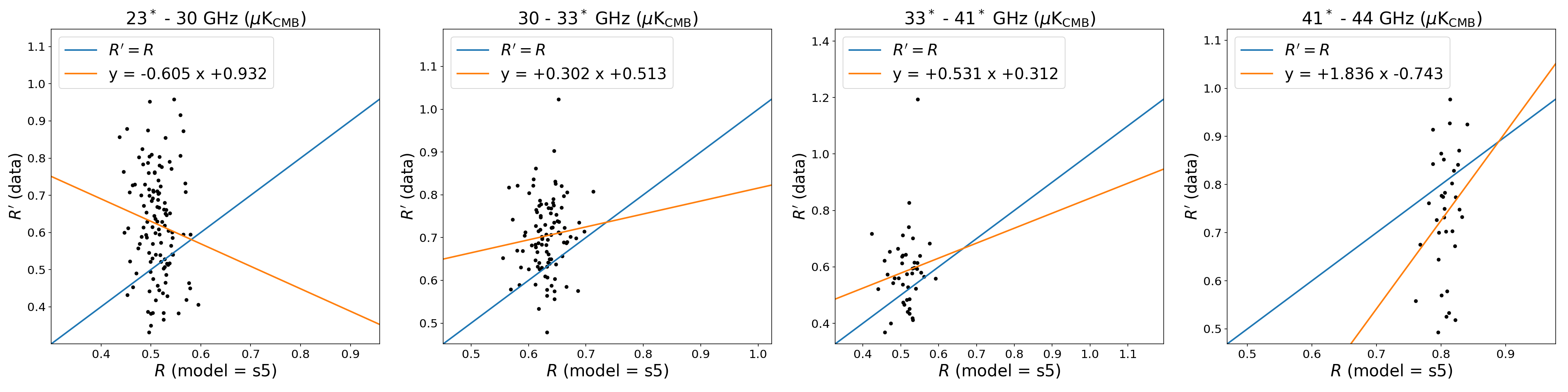}
\includegraphics[width=0.8\textwidth]{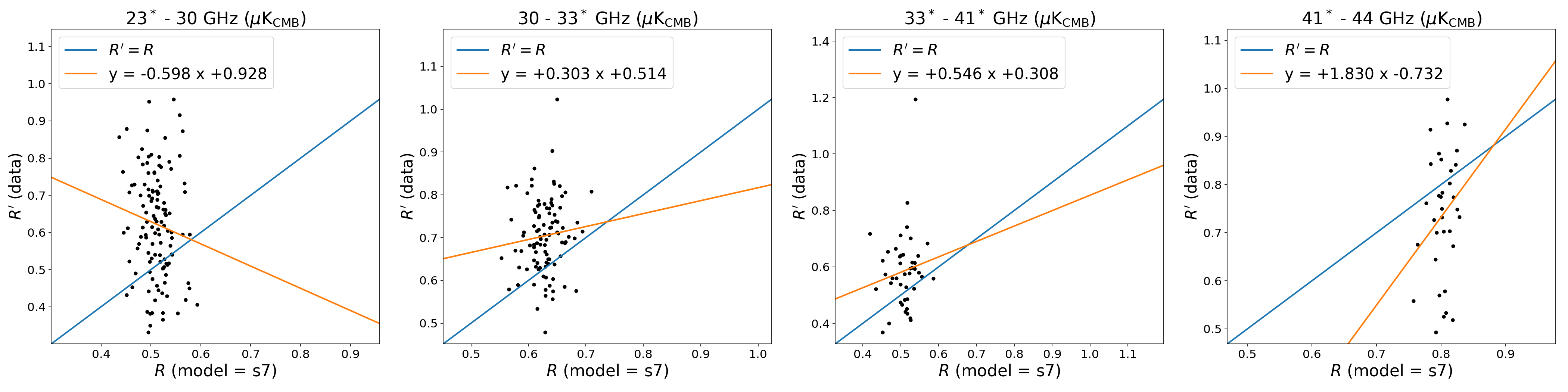}
\includegraphics[width=0.8\textwidth]{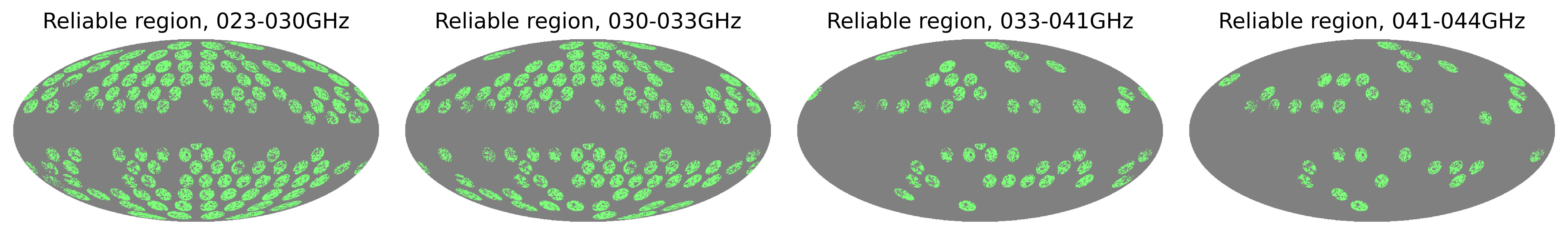}
\caption{Scatter points of $R^\prime$ versus $R$ for synchrotron in the reliable regions with $C^\prime \geq 0.80$ in green (bottom row). 
Frequency bands with star represent WMAP bands. }
\label{fig:scatter:synchrotron}
\end{figure*}
In this section, we test another important Galactic foreground, synchrotron radiation, which arises from high-energy cosmic-ray electrons undergoing relativistic gyration in interstellar magnetic field. 
It dominates the polarized sky emission below 70 GHz and also constitutes a major foreground component in temperature mode at these frequencies \citep{2013A&A...553A..96D}. 
Due to the energy distribution of Galactic cosmic-ray electrons, synchrotron emission can be approximated by a power-law spectrum with a spectral index of approximately $-1.11$ (in SI unit, \citealp{2016A&A...594A..10P}). 
The \textit{Planck} Collaboration proposed a simplified model for synchrotron, expressed as:
\begin{equation}
I_\mathrm{s}(\nu) = I_\mathrm{s}(\nu_0) \frac{f_\mathrm{s}(\nu)}{f_\mathrm{s}(\nu_0)},
\end{equation}
where $I_\mathrm{s}(\nu_0)$ is the synchrotron intensity map at $\nu_0 = 408\,\mathrm{MHz}$, based on \citet{1981A&A...100..209H, 1982A&AS...47....1H}, and $f_\mathrm{s}(\nu)$ represents the spectral energy distribution of synchrotron as adopted by the \textit{Planck} Collaboration. 
In this model, the spectral index is treated as a fixed and spatially uniform parameter across the sky. 
However, since the methodology employed in this study emphasizes the model's ability to fit data at local scales, we focus on synchrotron models characterized by spatially varying spectral indices. 
Specifically, we examine the synchrotron emission models implemented in the \texttt{PySM3} package, denoted as models s1, s2, s3, s5, and s7. 
Each of these models can be expressed in a unified form as a power-law spectrum: 
\begin{equation}
I_\mathrm{s}(\nu) = I_\mathrm{s}(\nu_0) \left(\dfrac{\nu}{\nu_0}\right)^{\beta_\mathrm{s}(\nu)},  
\end{equation}
where $I_\mathrm{s}(\nu_0)$ is the synchrotron intensity at reference frequency $\nu_0$, and $\beta_\mathrm{s}(\nu)$ is the spectral index, which may be constant or frequency-dependent.
Model s1 uses the reprocessed $408\,\mathrm{MHz}$ Haslam map \citep{1981A&A...100..209H, 1982A&AS...47....1H} from \citet{2015MNRAS.451.4311R} as the reference intensity map $I_\mathrm{s}(\nu_0)$, and adopts the spectral index map $\beta_\mathrm{s}$ from the model 4 of \citet{2008A&A...490.1093M}. 
Model s2 retains the same reference intensity map as s1, but simplifies the spatial variation of the spectral index by modeling it as a smooth function of Galactic latitude $b$: 
\begin{equation}
\beta_\mathrm{s} = -3.0 - 0.3 \sin\left|b\right|. 
\end{equation}
This formulation reflects the physical expectation that synchrotron at higher Galactic latitudes is produced by older, lower-energy cosmic-ray electrons that have diffused out of the Galactic plane \citep{2012ApJ...753..110K, 2014PTEP.2014fB109I}, resulting in a steeper synchrotron spectrum \citep{2007ApJ...665..355K, 2014ApJ...790..104F, 2015ApJ...811...89R}. 
Model s3 also uses the same $I_\mathrm{s}(\nu_0)$ map as s1, but introduces a frequency-dependent curved spectral index of the form: 
\begin{equation}
\beta_\mathrm{s}(\nu) = \beta_\mathrm{s}(\nu_c) + c_\mathrm{s} \ln\left(\dfrac{\nu}{\nu_c}\right), 
\end{equation}
where $c_\mathrm{s} = - 0.052$ according to \citet{2012ApJ...753..110K} and $\nu_c = 23\,\mathrm{GHz}$. 
The spectral index map at $\nu_c$, i.e., $\beta_\mathrm{s}(\nu_c)$, is taken from the model 1 of \citet{2008A&A...490.1093M}. 
Model s5 adopts a reference intensity map similar to that of s1, but transforms the $408\,\mathrm{MHz}$ template to $23\,\mathrm{GHz}$ using a fixed power-law spectrum with $\nu^{-1.1}$ (in SI unit), such that the reference frequency becomes $\nu_0 = 23\,\mathrm{GHz}$ \citep{2025arXiv250220452T}. 
Additionally, its spectral index map is updated using observations from the S-PASS survey \citep{2018A&A...618A.166K}, resulting in a refined spatial distribution of $\beta_\mathrm{s}$. 
Model s7 is also a curved power-law model. 
It utilizes the same $I_\mathrm{s}(\nu_0)$ and $\beta_\mathrm{s}$ maps as s5, but replaces the constant curvature parameter $c_\mathrm{s} = -0.052$ used in s3 with a spatially varying curvature map derived to match the results of the ARCADE experiment \citep{2012ApJ...753..110K}. 
All intensity and spectral index templates for these models are bundled within the \texttt{PySM3} data directory.\footnote{\url{https://portal.nersc.gov/project/cmb/pysm-data/}}

Figure \ref{fig:scatter:synchrotron} compares the performance of these five synchrotron models in the frequency range of 23---44 GHz. 
Due to the limited number of frequency channels available in \textit{Planck} LFI within this range, we complement the analysis with WMAP observations \citep{2013ApJS..208...20B} at K band (23 GHz), Ka band (33 GHz), and Q band (41 GHz). 
It should be noted that WMAP does not provide detailed spectral transmission, and therefore no color correction is applied to its maps. 
However, this omission has a negligible impact on the scatter distributions relevant to this study (see \citealp{2025ApJS..276...45L} for a quantitative assessment). 

The results shown in Figure \ref{fig:scatter:synchrotron} indicate that existing synchrotron models can capture basic frequency dependence of the emission at low frequency, but all models fail to describe the variation of the synchrotron intensity ratio across the sky. 
Therefore, none of these models is sufficiently accurate.

\section{Test of other foreground components}
\label{sec: test of other components}
Compared to the modeling of thermal dust and synchrotron emission, the physical characterization of free-free emission and AME remains less developed. 
In this section, we test the models of these latter two components. 
Although free-free emission and AME exhibit higher intensity in the 40--100 GHz band compared to other frequencies, this band suffers from significant blending of multiple foreground components (as well as the CMB), for which our modeling remains coarse. 
Consequently, the analysis of free-free emission and AME presented here is still primarily. 
\subsection{Free-free emission}

\begin{figure}[htbp]
\centering
\includegraphics[width=0.23\textwidth]{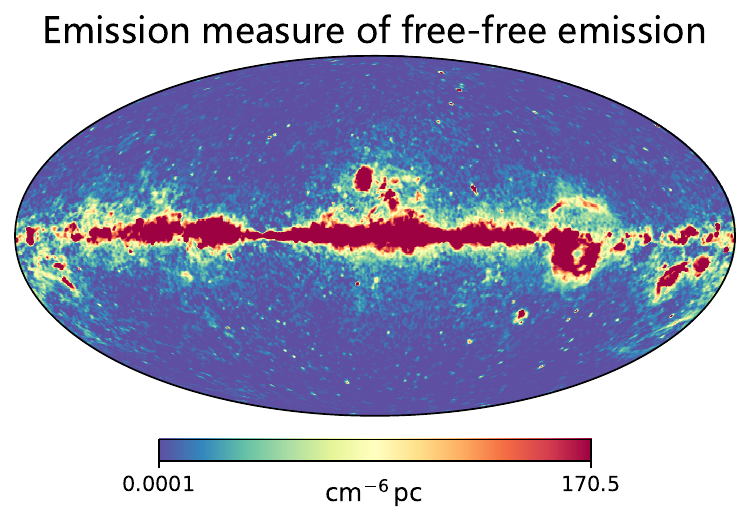}
\includegraphics[width=0.23\textwidth]{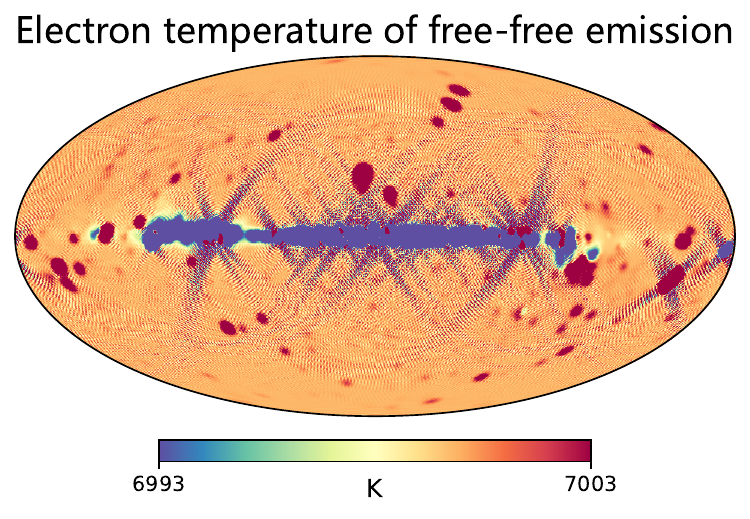}
\caption{Free-free emission parameters derived by 
\cite{2016A&A...594A..10P}. 
Left: Emission measure (EM) of free electrons. 
The modeling of free-free emission is significantly contaminated by compact sources, which introduces uncertainties in the derived EM map \citep{2016A&A...594A..10P, 2016A&A...594A..25P}.
Right: Electron temperature. 
The distribution has a median value of $7000\,\mathrm{K}$ with a standard deviation of only 93 K, roughly 1\%. }
\label{fig:free-free_EM_free-free_T}
\end{figure}

\begin{figure}[htpb]
\centering
\includegraphics[width=0.37\textwidth]{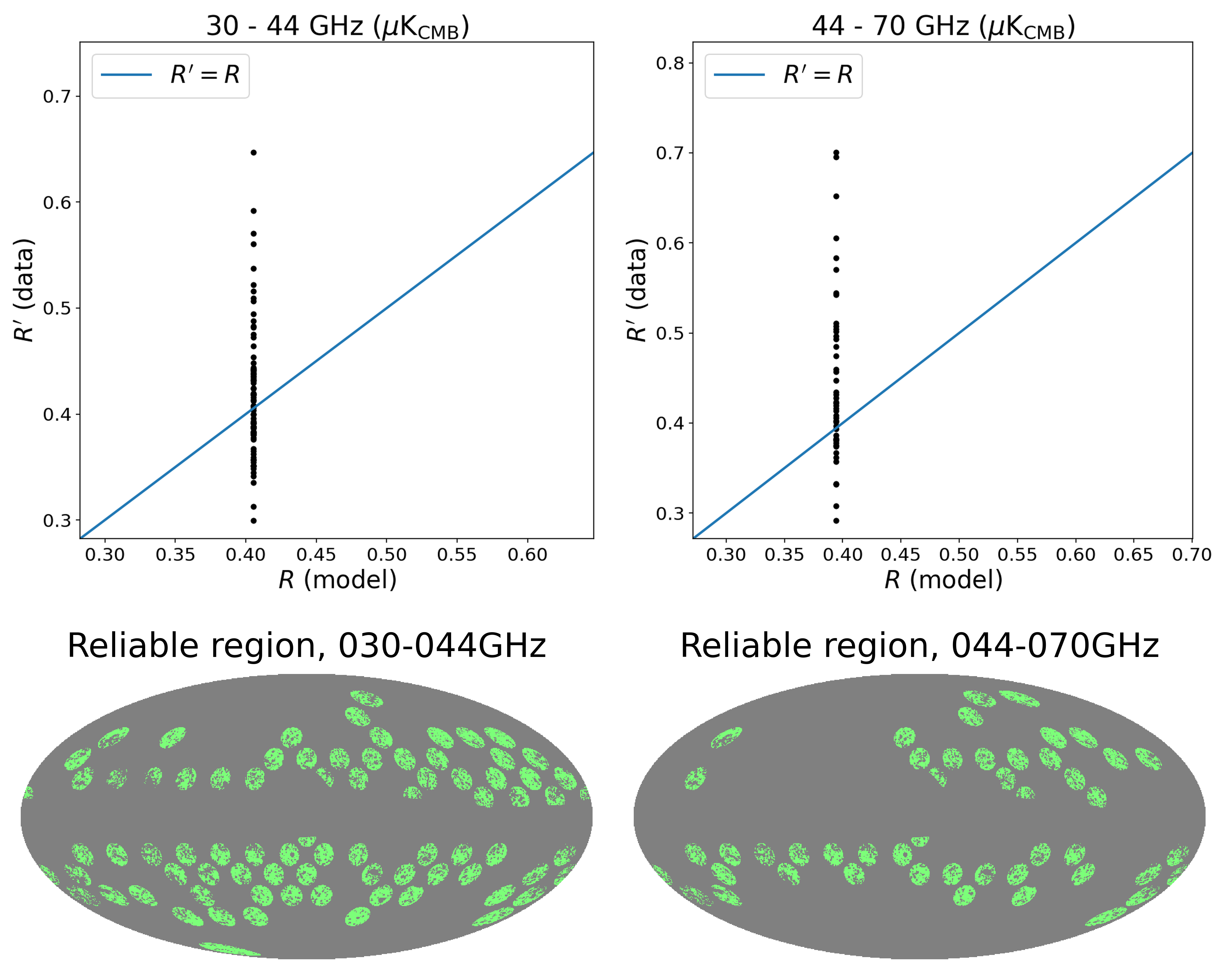}
\caption{Scatter points of $R^\prime$ versus $R$ for free-free emission (upper) in the reliable regions with $C^\prime \geq 0.80$ in green (lower). }
\label{fig:free-free:scatter}
\end{figure}

Free-free emission originates from Coulomb scattering between non-relativistic free electrons and atomic nuclei in warm ionized medium, 
constituting another major Galactic foreground component below 70 GHz. 
In this subsection, we evaluate the free-free emission model adopted by the \textit{Planck} Collaboration \citep{2016A&A...594A..10P}, which is based on the formulation provided by \citet{2011piim.book.....D}: 
\begin{equation}
I_{\mathrm{ff},\nu} = \big[1-\exp\left(-\tau_{\mathrm{ff},\nu}\right)\big] T_\mathrm{e}, 
\end{equation}
where $\tau_{\mathrm{ff},\nu}$ is the optical depth of warm ionized medium clouds: 
\begin{equation}
\tau_{\mathrm{ff},\nu} = 0.0546798 \,\nu_9^{-2} \left[T_\mathrm{e}\right]^{-3/2}g_\mathrm{ff}\,\left[\mathrm{EM}\right]. 
\end{equation}
Electron temperature $T_\mathrm{e}$ and emission measure $\mathrm{EM}$ are the two free parameters in this model 
and $g_\mathrm{ff}$ is the gaunt factor: 
\begin{equation}
g_\mathrm{ff} = \ln\left\{\exp\left[5.96023-\frac{\sqrt{3}}{\mathrm{\pi}}\ln\left(Z_\mathrm{i}\nu_9 T_4^{-3/2}\right)\right] + \mathrm{e}\right\}, 
\end{equation}
here $T_4 = T / \left(10^4\,\mathrm{K}\right)$, $\nu_9 = \nu / \mathrm{GHz}$. 

Figure \ref{fig:free-free_EM_free-free_T} presents the two free parameters, $T_\mathrm{e}$ and $\mathrm{EM}$,\footnote{\href{https://irsa.ipac.caltech.edu/data/Planck/release_2/all-sky-maps/previews/COM_CompMap_freefree-commander_0256_R2.00/index.html}{COM\_CompMap\_freefree-commander\_0256\_R2.00.fits}} in the \textit{Planck} free-free emission model. 
However, it is evident that $T_\mathrm{e}$ exhibits negligible spatial variation across the sky, effectively reducing the number of free parameters to only one. 
This simplification substantially limits the model's ability to capture the spatial complexity of free-free emission, as demonstrated in Figure \ref{fig:free-free:scatter}. 

\subsection{Anomalous Microwave Emission}

\begin{figure}[htpb]
\centering
\includegraphics[width=0.47\textwidth]{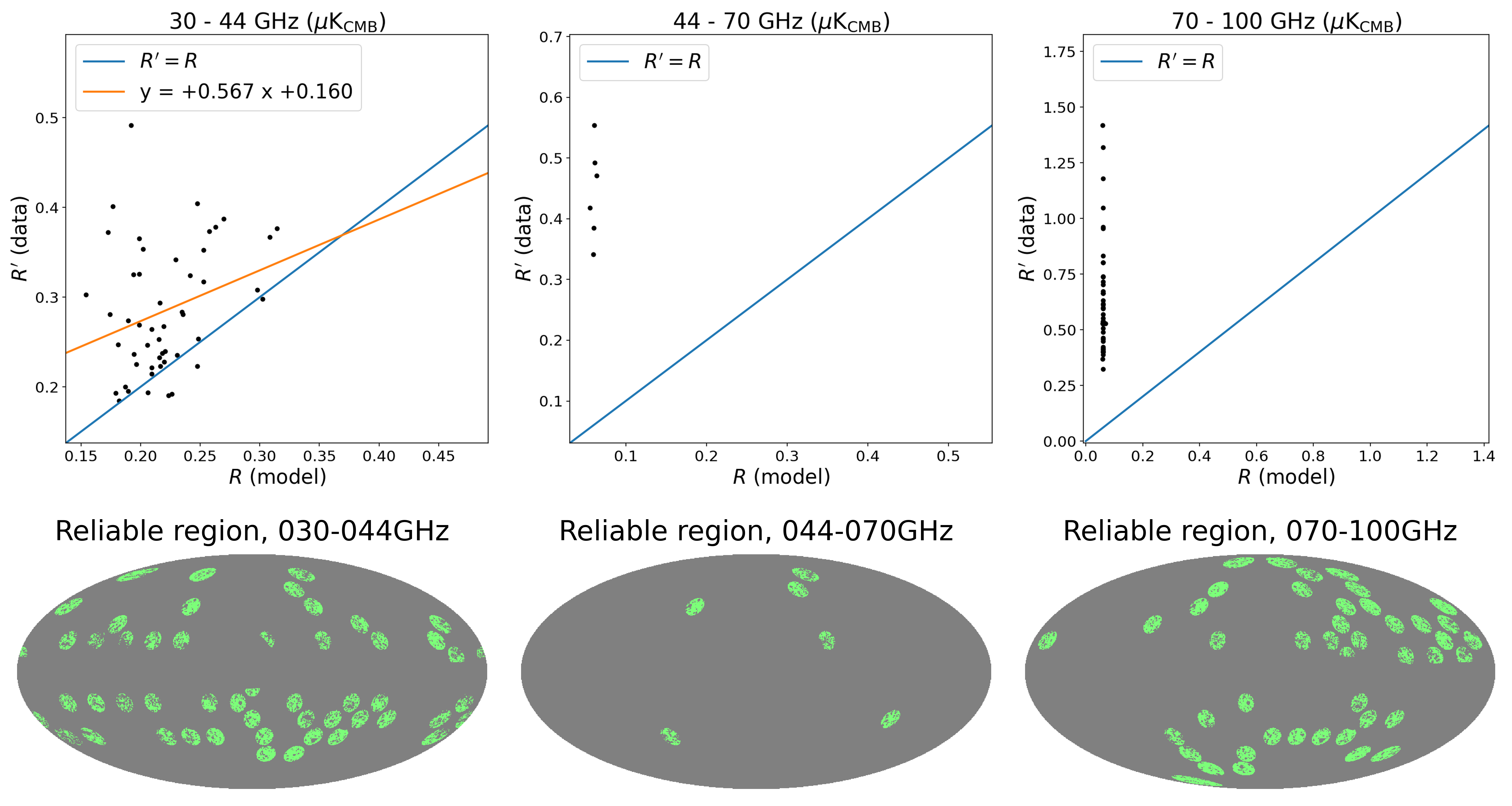}
\caption{Scatter points of $R^\prime$ versus $R$ for AME (upper) in the reliable regions with $C^\prime \geq 0.80$ in green (lower). }
\label{fig:AME:scatter}
\end{figure}

The aforementioned foreground components are insufficient to account for the observed results in the microwave band. 
\citet{1997ApJ...486L..23L} was the first to recognize the presence of a hidden component and referred to it using a term analogous to ``anomalous microwave emission.''
\citet{2016A&A...594A..10P} develop a model for AME based on calculations using the SpDust2 code\footnote{\url{https://ascl.net/1010.016}} \citep{2009MNRAS.395.1055A, 2011MNRAS.411.2750S}. 
This model decomposes AME into two distinct spinning-dust components: 
\begin{equation}
\begin{aligned}
I_\mathrm{sd}(\nu)
&= I^1_\mathrm{sd}(\nu_0)\frac{f_\mathrm{sd}\left(\dfrac{\nu_\mathrm{p0}}{\nu^1_p}\cdot\nu\right)}{f_\mathrm{sd}\left(\dfrac{\nu_\mathrm{p0}}{\nu^1_p}\cdot\nu^1_0\right)} + 
I^2_\mathrm{sd}(\nu_0)\frac{f_\mathrm{sd}\left(\dfrac{\nu_\mathrm{p0}}{\nu^2_p}\cdot\nu\right)}{f_\mathrm{sd}\left(\dfrac{\nu_\mathrm{p0}}{\nu^2_p}\cdot\nu^2_0\right)}, 
\end{aligned}
\end{equation}
where $\nu^1_0$ and $\nu^2_0$ are the reference frequency of two spinning-dust components: 
\begin{equation}
\begin{aligned}
\nu^1_0 &= 22.8\,\mathrm{GHz}, \\
\nu^2_0 &= 41.0\,\mathrm{GHz}. 
\end{aligned}
\end{equation}
In addition, $\nu_p^1$ and $\nu_p^2 = 33.35\,\mathrm{GHz}$ are peak frequency of these two components. 
\textit{Planck} AME model contains: 
three free parameters, $\nu^1_p$, $T^1_{\mathrm{sd}}(\nu_0) = \dfrac{c^2}{2k_\mathrm{B}^2\nu^2} I^1_{\mathrm{sd}}(\nu_0)$, $T^2_{\mathrm{sd}}(\nu_0) = \dfrac{c^2}{2k_\mathrm{B}^2\nu^2} I^1_{\mathrm{sd}}(\nu_0)$, 
one fixed parameter $\nu^2_p = 33.35\,\mathrm{GHz}$, 
and emission template of each component $f_\mathrm{sd}(\nu)$. 
All of them are recorded in \textit{Planck} .fits file,\footnote{\href{https://irsa.ipac.caltech.edu/data/Planck/release_2/all-sky-maps/previews/COM_CompMap_AME-commander_0256_R2.00/index.html}{COM\_CompMap\_AME-commander\_0256\_R2.00.fits}} 
where both $T^1_{0,\mathrm{sd}}$ and $T^2_{0,\mathrm{sd}}$ are in resolution with $\mathrm{FWHM} = 60^\prime$. 

The scatter plots in Figure \ref{fig:AME:scatter} reveal significant deviations in AME modeling within the 44---100 GHz range, indicating either inadequacies in capturing AME behavior or unidentified components in this frequency range. 
While the 30---44 GHz range appears better fit, this is likely attributable to noise-induced broadening along the x-axis rather than a true physical match.

\subsection{Residuals}
\begin{figure}[htpb]
\centering
\includegraphics[width=\linewidth]{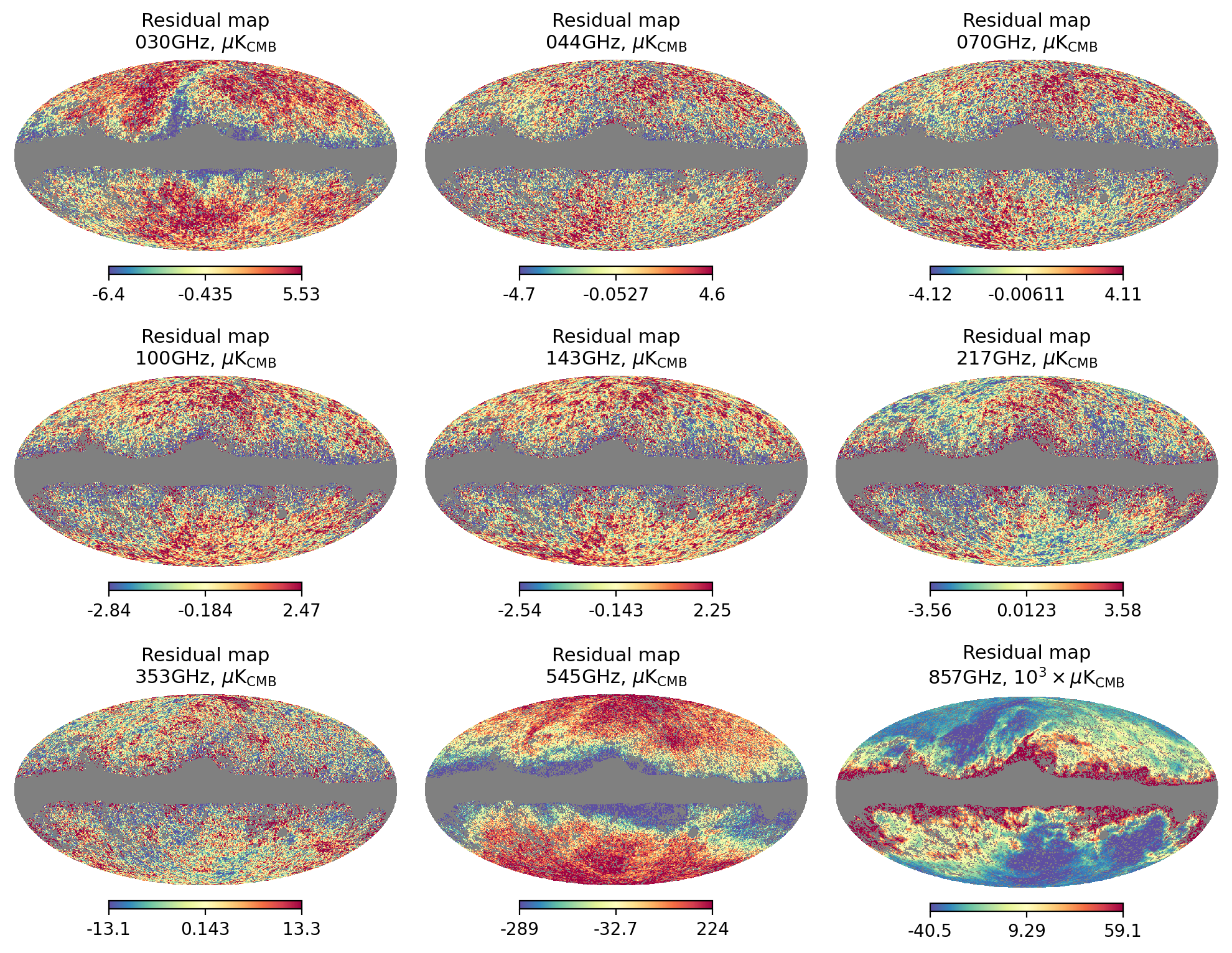}
\caption{Rows 1---3: Maps of residuals 
(monopoles and dipoles are excluded)
from 30 to 857 GHz. 
The gray regions are excluded from the statistical analysis in this study due to contamination from compact sources and complex Galactic foreground emission near the Galactic disk. 
Traces of zodiacal emission are marginally visible at 44, 70, 100 and 143 GHz bands. 
}
\label{fig:map:residual}
\end{figure}

To further assess the reliability of the data processing, Figure~\ref{fig:map:residual} presents the residuals across the nine \textit{Planck} frequency bands from 30 to 857 GHz. 
These residual maps are obtained by subtracting thermal dust emission (as modeled by model Planck15-C
\footnote{
The use of Planck15-C dust model here does not mean it is better than other dust models, because all dust models are suspicious according to our results. 
}
), synchrotron (based on \textit{Planck} official model), free-free emission, and AME from each single-frequency \textit{Planck} sky map. 
The dominant components in the residuals are their monopoles and dipoles; 
however, these largest-scale signals do not affect the correlation analysis conducted in this paper, because they contribute largely to the local offsets. 
Therefore, in Figure~\ref{fig:map:residual}, the monopoles and dipoles have been removed. Meanwhile,
in Appendix~\ref{appendix:supplement}, Tables~\ref{tab:RMS_dust} and \ref{tab:RMS_synch} quantify the magnitude of the residuals (after removal of the monopoles and dipoles) by calculating their RMS amplitude. 

\section{Conclusion and discussion}
\label{sec:conclusions}

This work evaluates the performance of current Galactic foreground emission estimates---including those for thermal dust, synchrotron, free-free, and AME---in a purely data-driven manner.
Scatter plots are used to compare the estimates with observational data across various sky patches, with tight constraints ensured by selecting only patches exhibiting strong cross-correlation between two bands.
If the estimate and data agreed, scatter points should distribute around and along $R=R'$; 
however, all tests exhibit significant deviations from this expectation, typically manifesting as either substantially different slopes or much narrower distributions along the horizontal axis.
While both over-simplified and over-complex models can degrade fitting performance, existing approaches predominantly adopt oversimplified assumptions (one component per emission mechanism, with the exception of the Meisner model for thermal dust). 
These discrepancies are most likely caused by the neglect of foreground complexity, either along or perpendicular to the line of sight. 
In particular, the latter is responsible for the narrowed distributions of scatter points.

Therefore, these discrepancies most likely stem from neglected foreground complexity, whether along or perpendicular to the line of sight. In particular, complexity perpendicular to the line of sight is the primary cause of the narrowed scatter distribution.

Results for specific foreground emission mechanisms are the following:
\begin{itemize}
\item \textbf{Thermal dust emission:} 
None of the estimates pass the test, though most perform reasonably well above 353 GHz. 
The primary issue is performance deterioration in 100---143 GHz, which is critical for cosmological inference. 
Among tested thermal dust models, \textit{Planck}'s single-component modified blackbody models (Planck13, Planck15-C, and Planck15-G) exhibit relatively better data consistency. 
This is likely due to their pixel-by-pixel dust parameter fitting, which is excellent at capturing the foreground complexity in different sky directions (perpendicular to the line of sight).

\item \textbf{Synchrotron emission:} 
Synchrotron emission estimates similarly fail in the test, compounded by increased component separation complexity in the 20---44 GHz range. 
Compared to the case of thermal dust, neglect of foreground complexity perpendicular to the line of sight (different sky directions) is more pronounced here, indicating a lower quality for synchrotron emission estimation since such problems can already be mitigated for thermal dust estimations.  

\item \textbf{Free-free emission:} \textit{Planck}'s free-free emission modeling assumes a nearly uniform electron temperature ($T_\mathrm{e} \sim 7000\,\mathrm{K}$). 
This simplification kills the model's ability to capture free-free emission's spatial variation.

\item \textbf{AME:} The poorest performance occurs for AME, which provides only crude results in the 30---44 GHz range.  
This deficiency likely arises from two factors: a relatively low amplitude and a incomplete physical understanding of AME.

\end{itemize}

The above results yield several key insights: 
\begin{enumerate}
\item Current understanding of Galactic foreground emissions remains inadequate, since existing estimates for all foreground emission mechanisms show significant discrepancies with observational data. 
While this clearly calls for additional frequency bands, the precise number required to resolve these discrepancies remains undetermined. 
However, we believe that deploying additional frequency bands always yields more substantial improvements than refining algorithms for foreground estimation.
  
\item Spatial constancy assumptions for key foreground parameters should be \emph{prohibited}, as all such simplifications generate substantially more artifacts than benefits. 

\item Straightforward approaches---particularly pixel-by-pixel estimations---lead to better results.  
\end{enumerate}

In summary, our findings highlight the critical importance of emphasizing foreground complexity and expanding the number of frequency bands in future CMB experiments. 
This emphasis is particularly timely in light of the challenges posed by inadequate modeling of Galactic foregrounds, which can significantly bias the search for primordial B-modes. 
For instance, as demonstrated in \citet{2025arXiv250800073L} based on analysis of the Simons Observatory, spatial variations in the dust frequency spectrum may lead to biases in the measurement of the tensor-to-scalar ratio. 
Thus, the issues highlighted above call for a careful consideration in the planning and design of next-generation CMB telescope projects: 
the required number of frequency bands, signal-to-noise ratio, and the consequent confidence level of foreground modeling; and a cost-benefit analysis of the two competing strategies---broader frequency coverage versus denser sampling across the spectrum.
We believe that a comprehensive solution for Galactic foreground emissions may require integrating all current and forthcoming CMB datasets---including WMAP, \textit{Planck}, LiteBIRD, PICO, and ground-based CMB-S4 observations:
Space missions are good at obtaining lower and higher frequency data, while ground missions can achieve superior signal-to-noise ratio at intermediate frequencies.
Thus, a joint analysis would be effective in addressing the challenges of complex Galactic foregrounds. 
The proposed methodology can also be applied to test models of polarized foreground components, as polarized observations benefit from expanded frequency coverage and enhanced data quality. 
In addition, it might also be necessary to consider space missions dedicated to the requirements of estimating Galactic foreground, as well as novel foreground estimation methods that are more powerful in handling foreground complexities.

\begin{acknowledgments}
This work is supported in part by National Key R\&D Program of China (2021YFC2203100, 2021YFC2203104), by NSFC (12433002, 12261131497), by CAS young interdisciplinary innovation team (JCTD-2022-20), by 111 Project (B23042), by Fundamental Research Funds for Central Universities, by CSC Innovation Talent Funds, by USTC Fellowship for International Cooperation, by USTC Research Funds of the Double First-Class Initiative, by the Anhui Provincial Natural Science Foundation 2308085MA30, and by the Anhui project Z010118169. This study utilizes observational maps from the latest Planck release~\citep{2020A&A...641A...3P}, as well as products regarding the mentioned Galactic foreground emission mechanisms.
\end{acknowledgments}

\bibliography{1_main}{}
\bibliographystyle{aasjournal}

\appendix

\section{Brief review of the method}
\label{app:method}

This section summarizes the methodology and statistic steps employed in this paper. 
Compared to \citet{2025ApJS..276...45L}, this paper implements minor refinements in the data processing methodology. 
While these refinements result in slight variations in the fitted regression lines for certain scatter plots relative to \citet{2025ApJS..276...45L}, they do not materially affect the final conclusions or compromise the overall validity of the results. 
\paragraph{Step 1: Foreground data maps}
Remove all components other than the target foreground component under examination
--- including CMB anisotropies, thermal dust emission, synchrotron, free-free emission, and AME---
from the observed single-frequency maps to obtain the foreground maps based on real observed data.\footnote{The carbon monoxide (CO) emission line is an exception; see Step 4. }
Due to the absence of totally reliable models for thermal dust and synchrotron emission, we adopt model Planck15-C for thermal dust removal, and the synchrotron model based on a spatially uniform spectral index, as employed by the \textit{Planck} Collaboration. 
Notably, the synchrotron template derived from the Haslam map \citep{1981A&A...100..209H, 1982A&AS...47....1H} contains residual monopole and dipole components. 
To address this, we follow the procedure of \citet{2017A&A...597A.131W}, which removes these large-scale residuals from the Haslam template prior to component separation.\footnote{$\text{Monopole} = 8.9\,\mathrm{K_{RJ}}$, and $\text{Dipole} = (3.2\,\mathrm{K_{RJ}}, 0.7\,\mathrm{K_{RJ}}, -0.8\,\mathrm{K_{RJ}})$ at 408 MHz. }
Given that \textit{Planck} sky maps adopt different unit conventions across frequency bands: 
the 30---353 GHz maps in CMB unit $\mathrm{K_{CMB}}$, while the 545---857 GHz maps in the radio unit $\mathrm{MJy\,sr^{-1}}$, 
to facilitate the calculation of linear regression ratios between adjacent frequency bands in analyses, we standardize all data-based foreground maps into the unit of $\mu\mathrm{K_{CMB}}$. 
For the maps at 545 and 857 GHz, the following unit conversion is performed: 
\begin{equation}
s_\mathrm{CMB} = s_\mathrm{SI}\, \dfrac{\displaystyle\int \left(\dfrac{\nu}{\nu_0}\right)^{-1} \mathcal{T}(\nu) \,\dd\nu}{\displaystyle\int b^\prime_\mathrm{CMB}(\nu)\mathcal{T}(\nu) \,\dd\nu}, 
\end{equation}
where $s_\mathrm{CMB}$ and $s_\mathrm{SI}$ denote the detector readings in units of $\mu\mathrm{K_{CMB}}$ and $\mathrm{MJy\,sr^{-1}}$, respectively, 
while $\mathcal{T}(\nu)$ and $\nu_0$ are the \textit{Planck} spectral transmission and the effective frequency for each channel, and 
\begin{equation}
b^\prime_\mathrm{CMB}(\nu) \equiv \dfrac{\partial B_\nu(T_0)}{\partial T}
\end{equation}
is the partial derivative of the Planck function with respect to temperature, evaluated at the mean temperature of the CMB, $T_0 = 2.7255\,\mathrm{K}$. 

\paragraph{Step 2: Foreground model maps}
Generate foreground model maps for these frequency bands based on their models. 
Color corrections are applied during the computation based on the spectral response of the \textit{Planck} detectors. 
\paragraph{Step 3: Smoothing}
Smooth the foreground component maps based on data and based on models with smoothing angle of $2^\circ$, to eliminate the influence from noises such as CIB anisotropies. 
\paragraph{Step 4: Mosaic disks}
Divide the maps into 
independent, non-overlapping
, equally sized mosaic disks with radius of $6^\circ$. 
These mosaic disks are defined by the Healpix resolution Nside=4. 
In each mosaic disk, calculate the correlation coefficient $C^\prime$ between foreground data maps in adjacent frequency bands. 
Select mosaic disks with $C^\prime \geqslant 0.95$ (for thermal dust) or $C^\prime \geqslant 0.80$ (for other components) as reliable regions. 
In this step, Galactic plane and compact sources have been masked. 
Due to the considerable uncertainty associated with CO emission lines, we mask regions where $\mathrm{CO}(2-1) > 1 \,\mathrm{K_{RJ}\,km\,s^{-1}}$ and exclude them from subsequent statistical analyses. 

\paragraph{Step 5: Linear regression ratio}
In the reliable regions, define the linear regression ratio between adjacent frequency bands within a mosaic disk as:  
\begin{equation}
\label{equ:R and R'}
\begin{dcases}
R = \dfrac{\sum_p(x_p-\overline{x})(y_p-\overline{y})}{\sum_p(x_p-\overline{x})^2}, &\text{model; }\\
R^\prime = \dfrac{\sum_{p}(x_p^\prime-\overline{x^\prime})(y_p^\prime-\overline{y^\prime})}{\sum_p(x_p^\prime-\overline{x^\prime})^2}, &\text{data. } 
\end{dcases}
\end{equation}
where $x_p$ and $y_p$ represent the emission intensities of the target foreground component at pixel $p$ in two adjacent frequency bands, respectively; 
$\overline{x}$ and $\overline{y}$ denote the average intensities within the same mosaic disk. 
Quantities with a prime denote foreground data maps, while quantities without a prime denote the foreground model maps. 
The linear regression ratio is fundamentally the slope of the least-squares regression line between paired sample sets $x$ and $y$ (or $x^\prime$ and $y^\prime$) from adjacent frequency bands. 
This quantity exhibits statistical robustness against overall sky-map offsets.

\paragraph{Step 6: Analysis in scatter plots}
Plot the scatter points of $R^\prime$ versus $R$ for adjacent frequency bands. 
If the scatters distribute closely with the expectation line $R' = R$, it indicates that the model provides a good fit to the real foreground emission. 
Conversely, significant deviations from this line suggest that the model performs poorly. 

Following the above procedures, $C^\prime \approx 1$ in a statistical region indicates that the residuals of non-target components in the foreground data map (obtained in Step 1) are significantly smaller than the target component's intensity within that region. 
Under the null hypothesis that the model adequately characterizes the target foreground component, the constraint $C^\prime \approx 1$ further implies that the model-derived linear regression ratio $R$ closely matches the data-derived ratio $R^\prime$. 
This encapsulates our core methodology: When $R$ and $R^\prime$ show substantial deviations in specific frequency bands, we reject the null hypothesis. 
\citet{2025ApJS..276...45L} provides rigorous validation via mathematical proofs and numerical simulations, establishing constraints on the regression line in the $R^\prime$–$R$ scatter plot under the assumption that the model provides a good fit to the data; see Figures 9 and 18 in \citet{2025ApJS..276...45L}. 
This methodology provides a robust statistical framework for evaluating the performance of Galactic foreground models across different frequency ranges. 

The raw data associated with this study are publicly available for download from the IPAC archive \citep{2020ipac.data.I473P, 2020ipac.data.I457P, 2020ipac.data.I460P, 2020ipac.data.I444P, 2020ipac.data.I463P, 2020ipac.data.I454P, 2020ipac.data.I469P, 2020ipac.data.I450P, 2020ipac.data.I465P, 2020ipac.data.I558P, 2020ipac.data.I559P}, 
from CDS archive \citep{2019yCat..36230021I}, 
from \url{http://sroll20.ias.u-psud.fr} \citep{2021A&A...650A..82D}, 
from \url{http://planck.skymaps.info} \citep{2014ascl.soft11012M}
. 
The pysm package \citep{2017ascl.soft04007T} used in this work is available on Zenodo \citep{andrea_zonca_2025_17254397}. 

\section{Supplementary Tables and Figures}
\label{appendix:supplement}
This section presents supplementary results beyond the scatter plots, including 
tables of residual RMS (Tables~\ref{tab:RMS_dust} and \ref{tab:RMS_synch}), 
intensity maps of thermal dust emission, synchrotron, free-free emission, and AME derived from satellite observations and foreground models (Figures~\ref{fig:intensity_maps_thermal dust}---\ref{fig:intensity_maps_AME}). 
Additionally, we show linear ratio maps of thermal dust and synchrotron emissions in Figures \ref{fig:maps_R_thermal dust}---\ref{fig:maps_R_synchrotron}. 

\begin{table}[H]
\centering
\caption{RMS of the residuals for different thermal dust models. }
\begin{threeparttable}
\centering
\begin{tabular*}{0.72\textwidth}{lcccccc} 
\toprule
 & 100 GHz & 143 GHz & 217 GHz & 353 GHz & 545 GHz & 857 GHz \\
\midrule
model Planck13   & $2.72$ & $3.00$ & $6.97$ & $10.2$ & $520$ & $1.70\times10^4$ \\
model Planck15-G & $2.76$ & $2.82$ & $5.77$ & $20.3$ & $402$ & $1.78\times10^4$ \\
model Planck15-C & $2.59$ & $2.11$ & $3.43$ & $11.2$ & $198$ & $4.19\times10^4$ \\
model Irfan19    & $2.39$ & $2.10$ & $3.77$ & $11.5$ & $485$ & $1.22\times10^4$ \\
model Meisner    & $1.94$ & $1.84$ & $3.43$ & $15.9$ & $261$ & $3.23\times10^4$ \\
model SRoll      & $2.54$ & $2.36$ & $3.16$ & $17.2$ & $402$ & $5.27\times10^4$ \\
model HD17-d5    & $3.85$ & $5.26$ & $9.98$ & $39.9$ & $183$ & $7.58\times10^4$ \\
model HD17-d7    & $3.82$ & $4.86$ & $8.41$ & $32.4$ & $195$ & $5.61\times10^4$ \\
model 3D         & $3.48$ & $4.53$ & $7.22$ & $23.9$ & $1.11\times10^3$ & $1.42\times10^5$ \\
\bottomrule
\end{tabular*}
\begin{tablenotes}
\item 
The residuals are obtained by subtracting the CMB anisotropies, AME, free-free emission, thermal dust emission (based on the various dust models tested in this paper), and synchrotron emission (using the \textit{Planck} official model) from the single-frequency \textit{Planck} sky maps, after which the monopole and dipole are removed from the remaining components. 
For each mosaic disk (see the bottom row of Figure~\ref{fig:scatter_3D}), the RMS of the residuals in single mosaic disk is computed. 
The pixel-number-weighted RMS of these values is then calculated to derive the overall residual RMS, which is reported in this table. 
All values in this table are in unit of $\mathrm{\mu K_{CMB}}$. 
\end{tablenotes}
\end{threeparttable}
\label{tab:RMS_dust}

\vspace{2cm}

\centering
\caption{RMS of the residuals for different synchrotron emission models. }
\begin{threeparttable}
\centering
\begin{tabular*}{0.49\textwidth}{lcccccc} 
\toprule
 & 23 GHz & 30 GHz & 33 GHz & 41 GHz & 44 GHz \\
\midrule
model s1 & 22.8 & 14.1 & 10.7 & 7.92 & 6.84 \\
model s2 & 19.9 & 13.1 & 10.2 & 6.97 & 6.24 \\
model s3 & 22.8 & 13.4 & 9.96 & 7.31 & 6.12 \\
model s5 & 10.8 & 6.41 & 6.86 & 5.88 & 4.69 \\
model s7 & 10.8 & 6.40 & 6.86 & 5.89 & 4.68 \\
\bottomrule
\end{tabular*}
\begin{tablenotes}
\item
Similar to Table~\ref{tab:RMS_dust}, but with the dust component subtracted based on the Planck15-C model, and synchrotron emission subtracted based on the synchrotron models tested in this paper. 
\end{tablenotes}
\end{threeparttable}
\label{tab:RMS_synch}
\end{table}

\begin{figure}[H]
\centering
\begin{minipage}{\textwidth}
\centering
\includegraphics[width=0.9\textwidth]{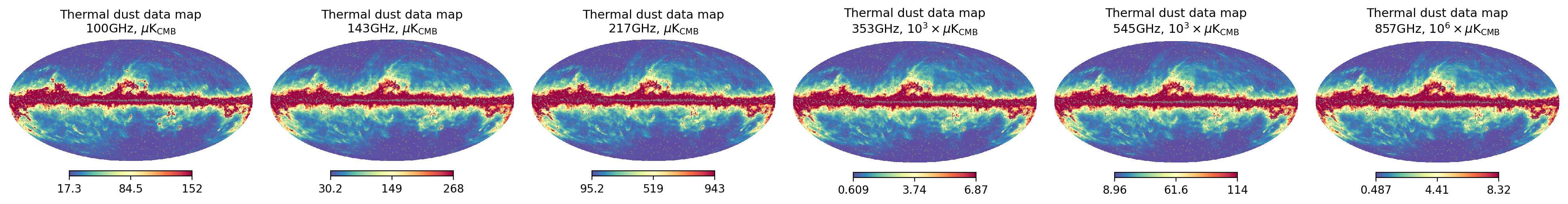}
\includegraphics[width=0.9\textwidth]{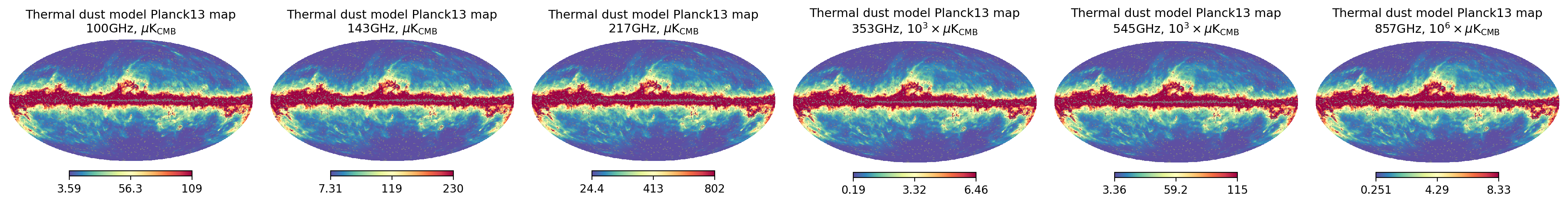}
\end{minipage}
\caption{(a) Upper section}
\end{figure}
\begin{figure}[H]
\centering
\begin{minipage}{\textwidth}
\centering
\includegraphics[width=0.9\textwidth]{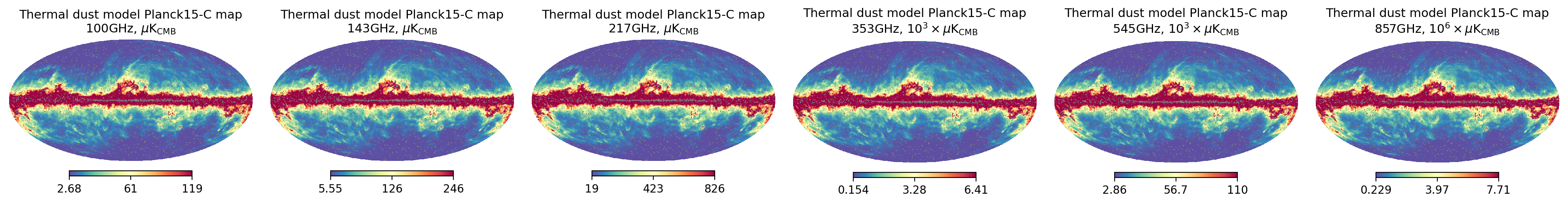}
\includegraphics[width=0.9\textwidth]{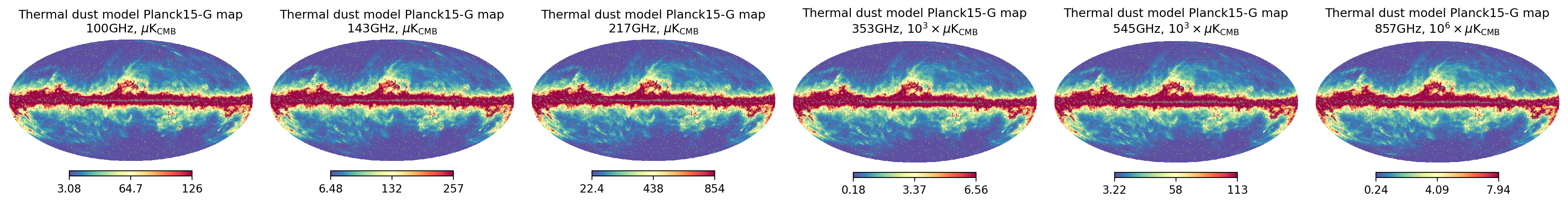}
\includegraphics[width=0.9\textwidth]{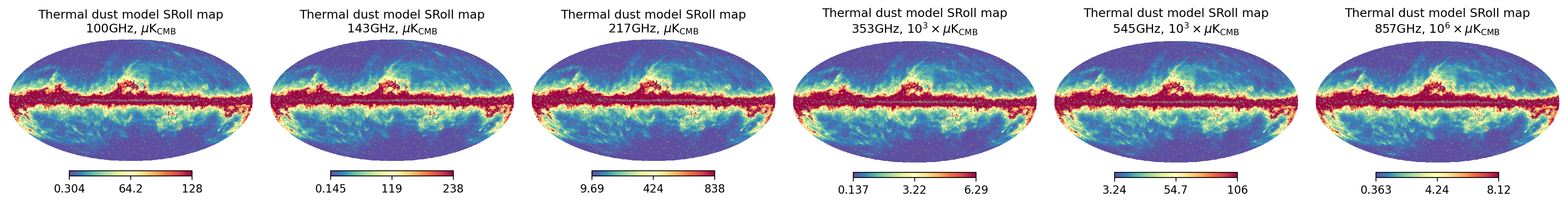}
\includegraphics[width=0.9\textwidth]{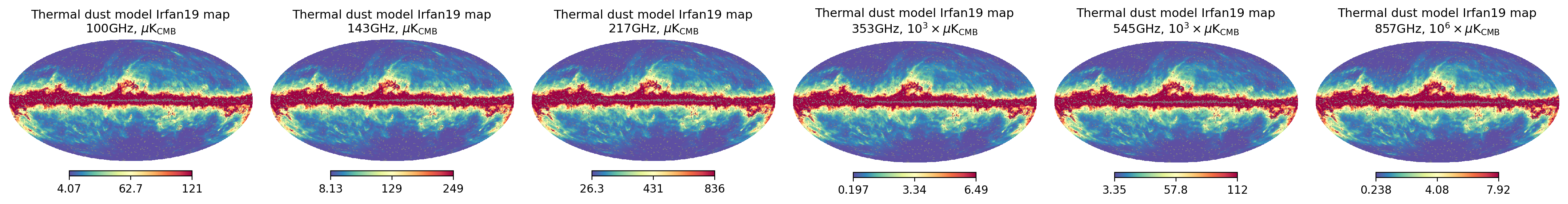}
\includegraphics[width=0.9\textwidth]{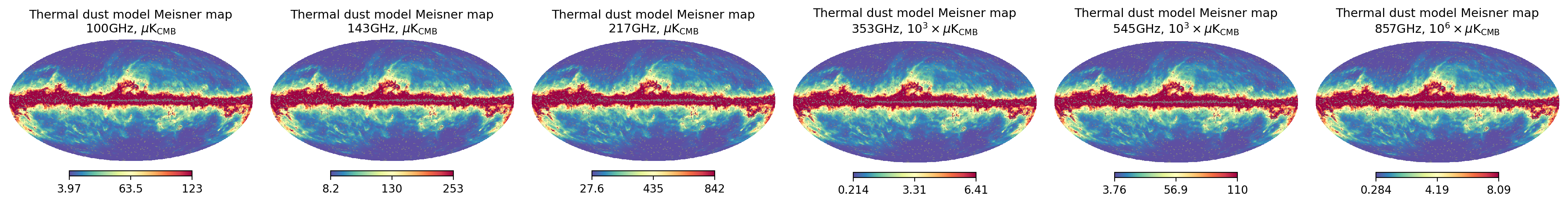}
\includegraphics[width=0.9\textwidth]{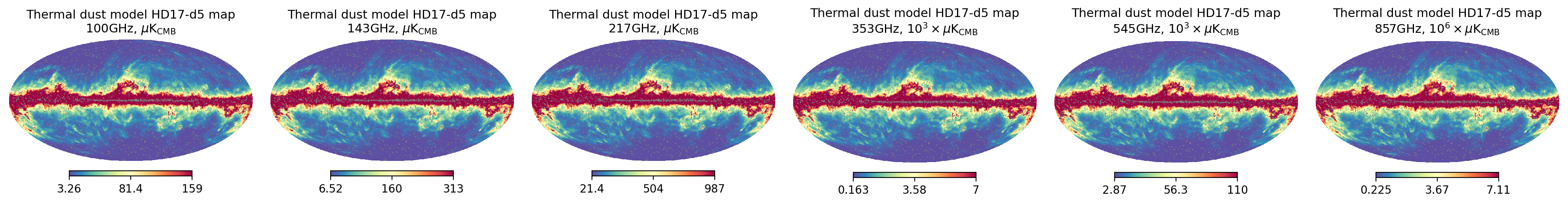}
\includegraphics[width=0.9\textwidth]{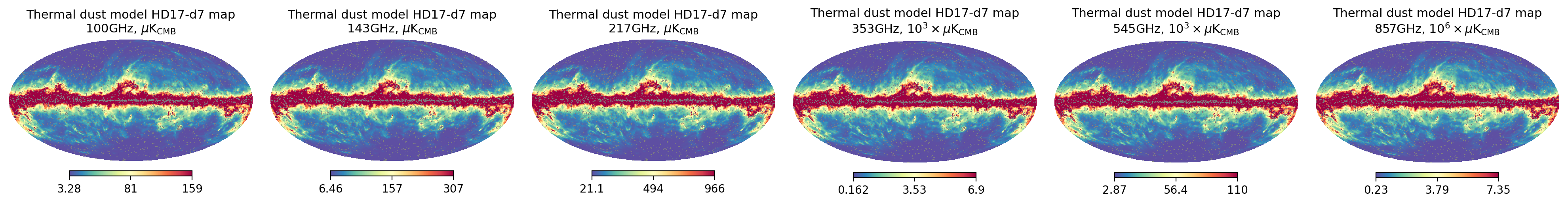}
\includegraphics[width=0.9\textwidth]{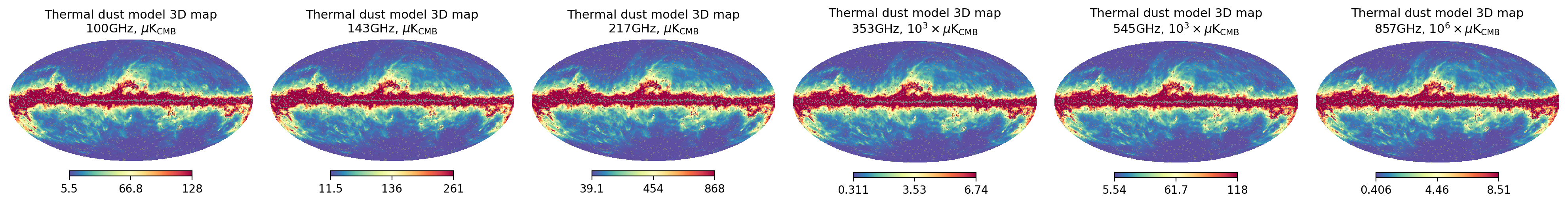}
\end{minipage}
\addtocounter{figure}{-1}
\caption{(b) Lower section}
\addtocounter{figure}{-1}
\caption{
Maps of thermal dust intensity based on observed data (row 1) and thermal dust emission models (rows 2---10). 
From left to right: 100, 143, 217, 353, 545, and 857 GHz. 
The color bar limits correspond to the 10th and 90th percentiles of the intensity distribution. 
}
\label{fig:intensity_maps_thermal dust}
\end{figure}

\begin{figure}[H]
\centering
\includegraphics[width=0.75\textwidth]{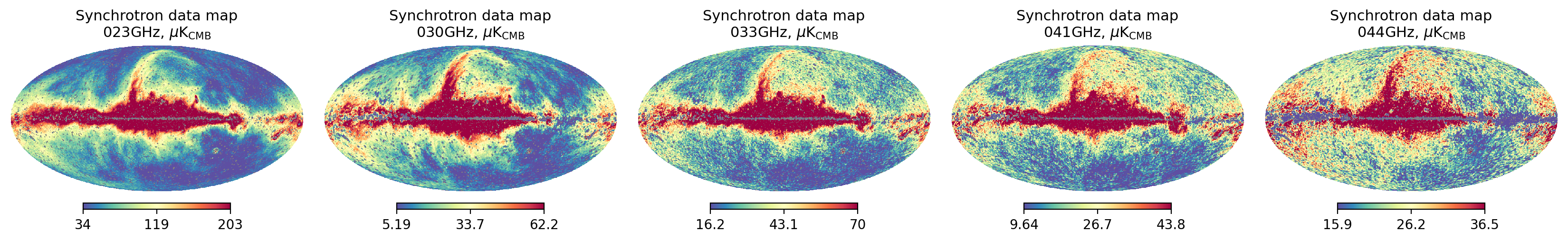}
\includegraphics[width=0.75\textwidth]{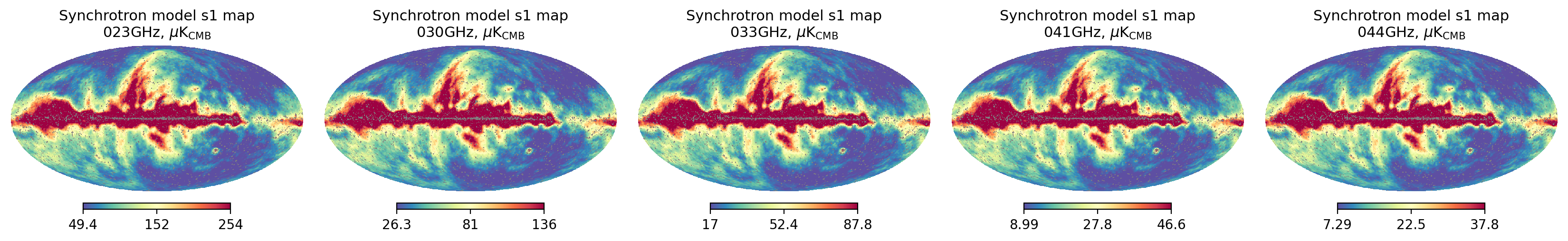}
\includegraphics[width=0.75\textwidth]{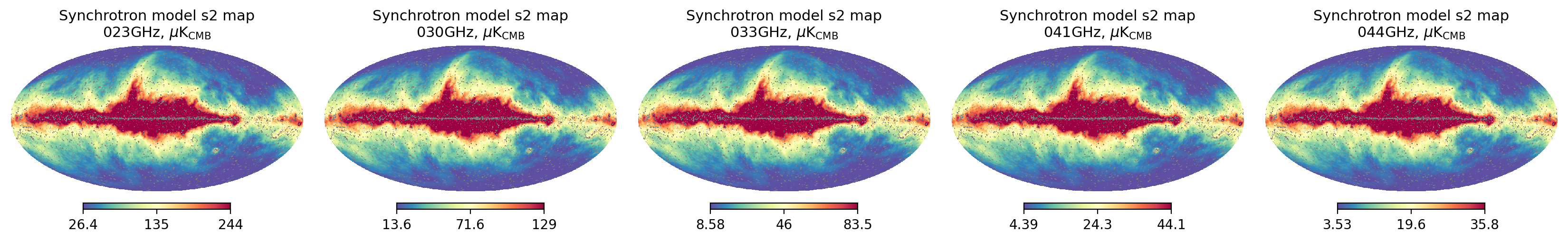}
\includegraphics[width=0.75\textwidth]{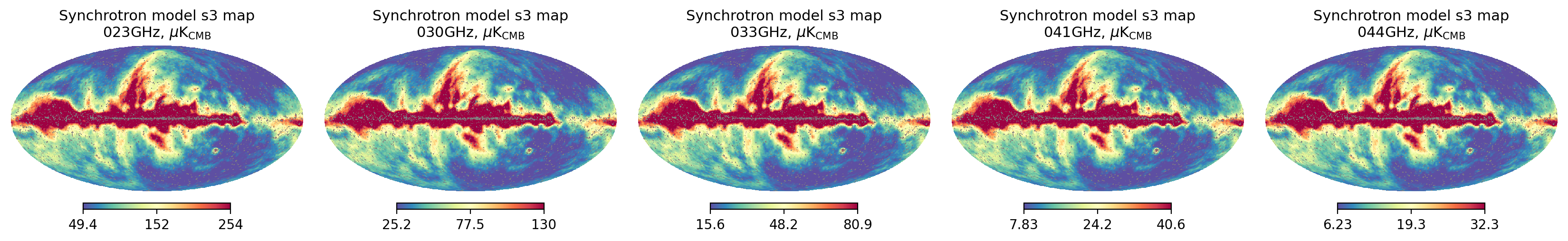}
\includegraphics[width=0.75\textwidth]{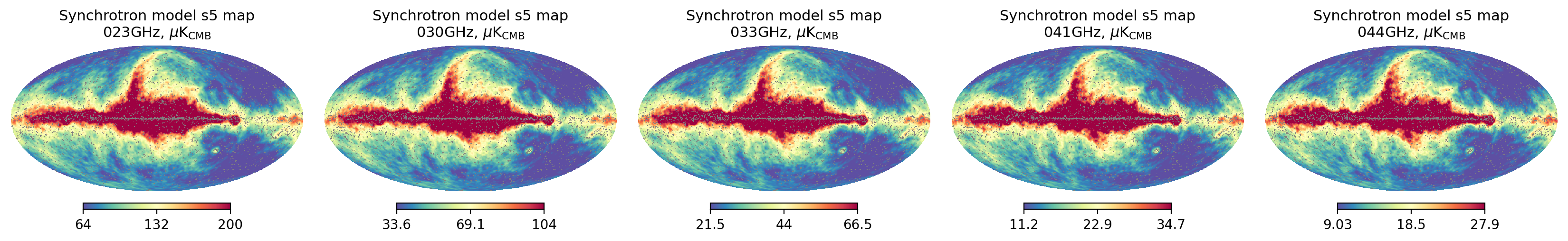}
\includegraphics[width=0.75\textwidth]{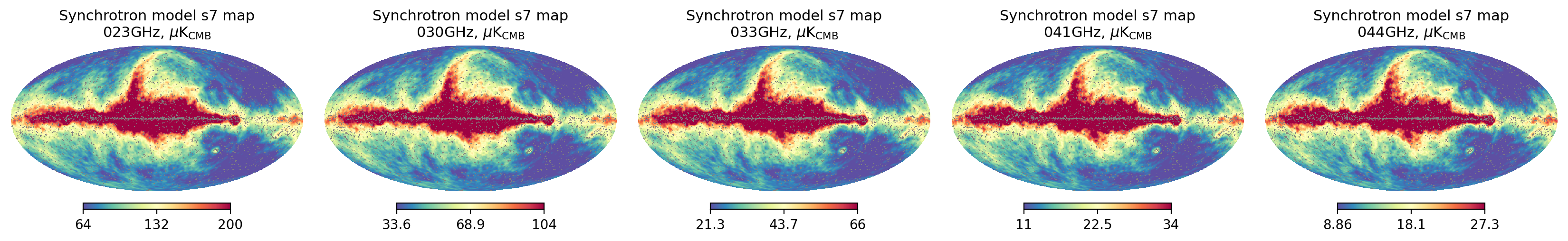}
\caption{
Maps of synchrotron intensity based on observed data (row 1) and synchrotron models (rows 2---6). 
}
\label{fig:intensity_maps_synchrotron}
\end{figure}

\begin{figure}[htpb]
\centering
\includegraphics[width=0.45\textwidth]{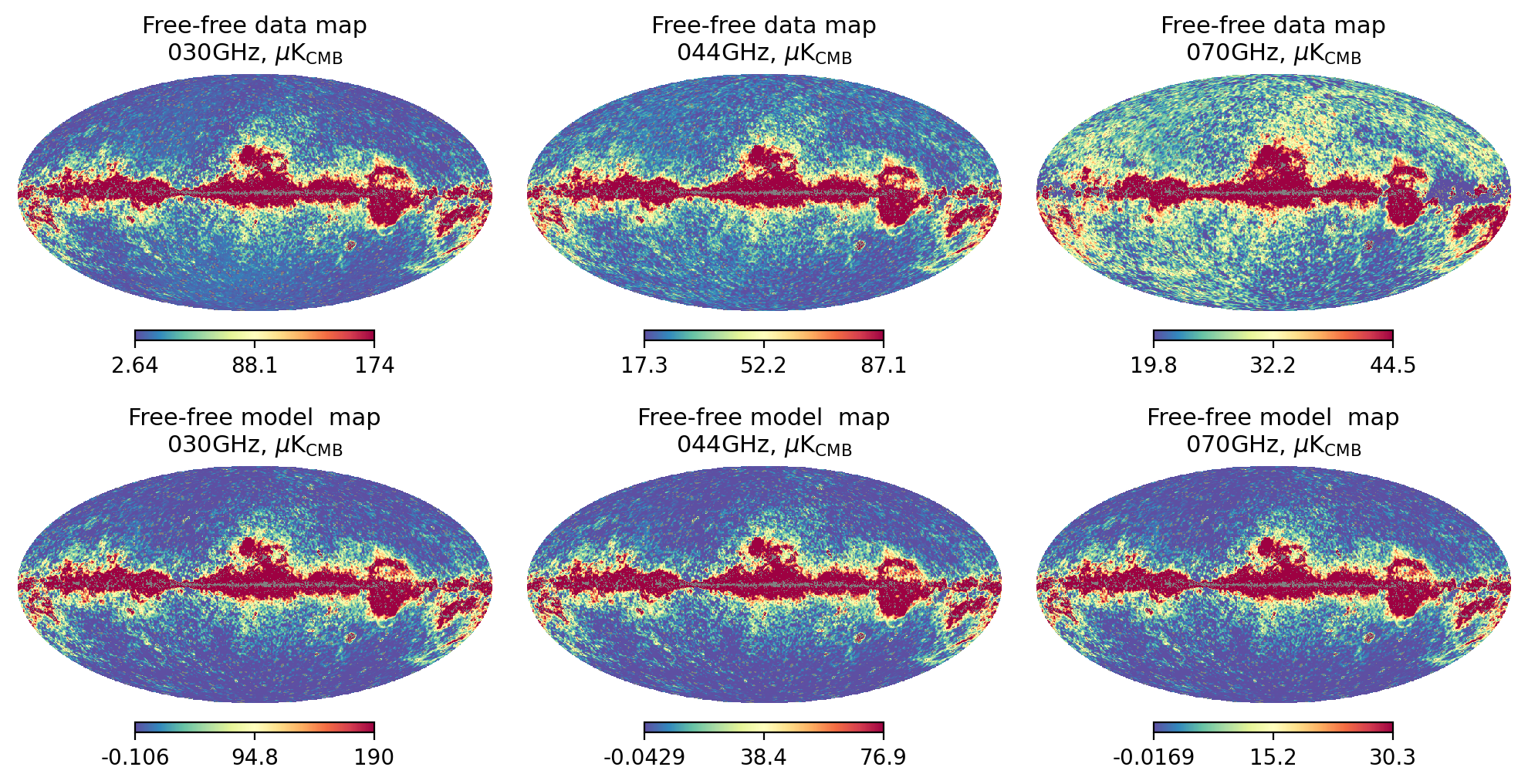}
\caption{Free-free intensity maps at 30, 44, and 70 GHz. 
All of data maps exhibit contamination from zodiacal light residuals. 
The free-free model maps exhibit minor negative values, artifacts resulting from Gibbs phenomenon during the sky map resolution adjustment (smoothing procedure).}
\label{fig:intensity_maps_free-free}
\end{figure}

\begin{figure}[htpb]
\centering
\includegraphics[width=0.45\textwidth]{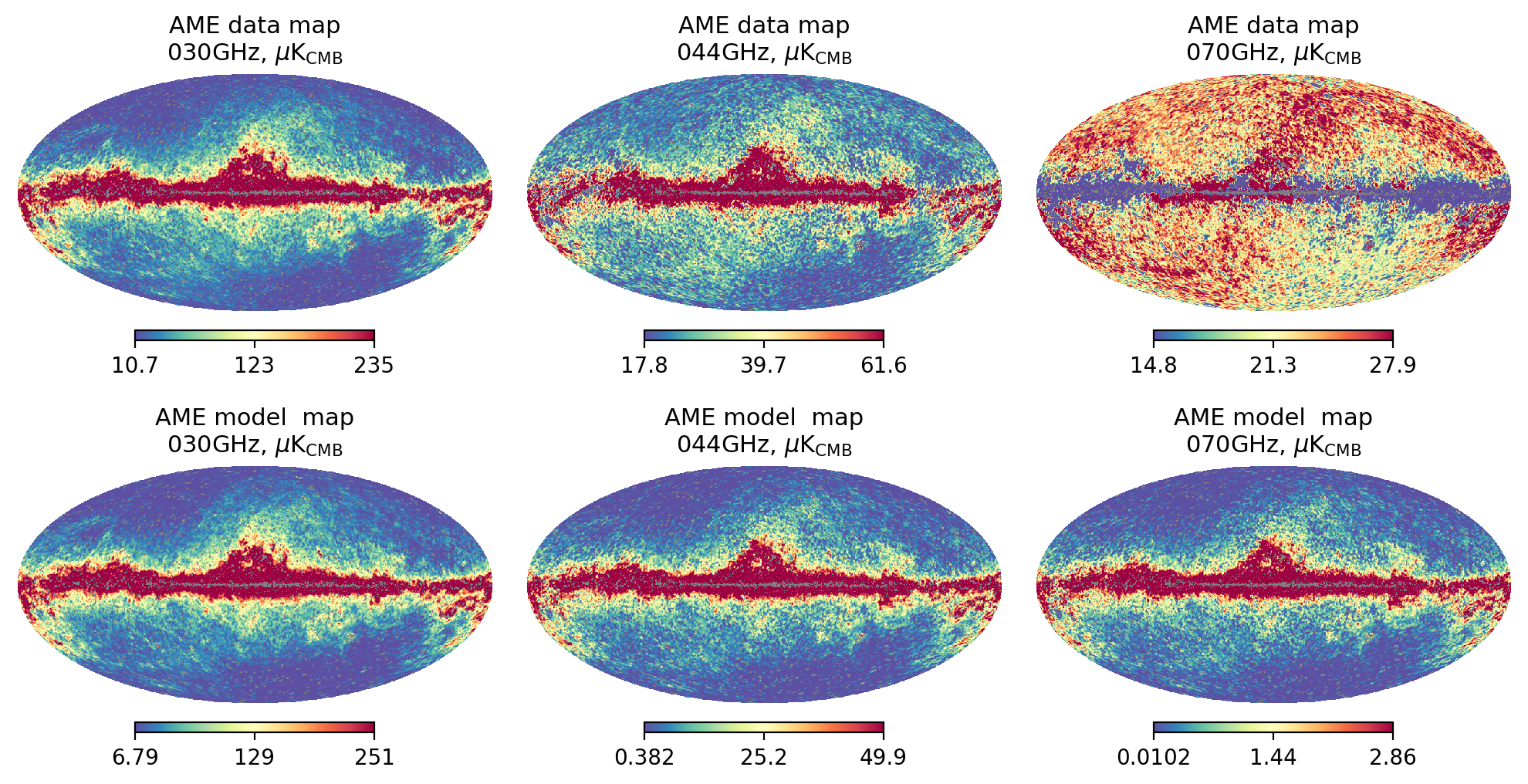}
\caption{AME intensity maps at 30, 44, and 70 GHz. 
Although \citet{2020A&A...641A...3P} claim they have completely subtracted zodiacal emission in their 2018 data release, 
residual zodiacal light contamination remains visually identifiable in the 70 GHz AME data map. 
}
\label{fig:intensity_maps_AME}
\end{figure}

\begin{figure}[htpb]
\centering
\includegraphics[width=0.75\textwidth]{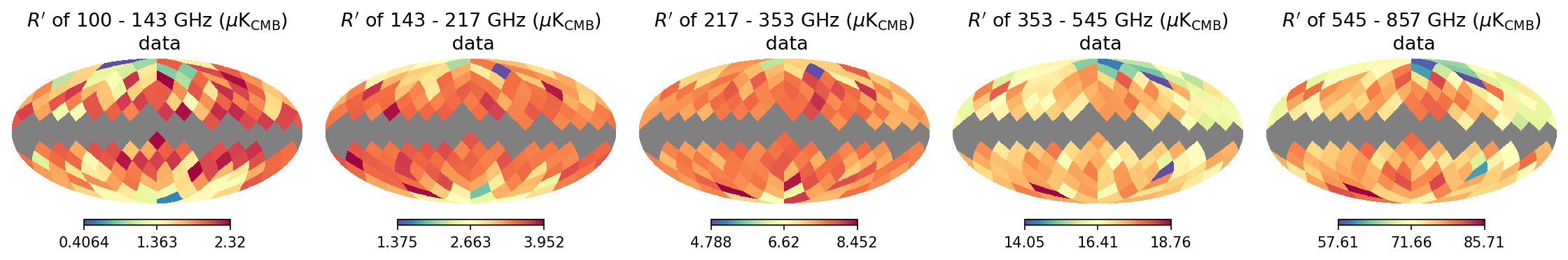}
\includegraphics[width=0.75\textwidth]{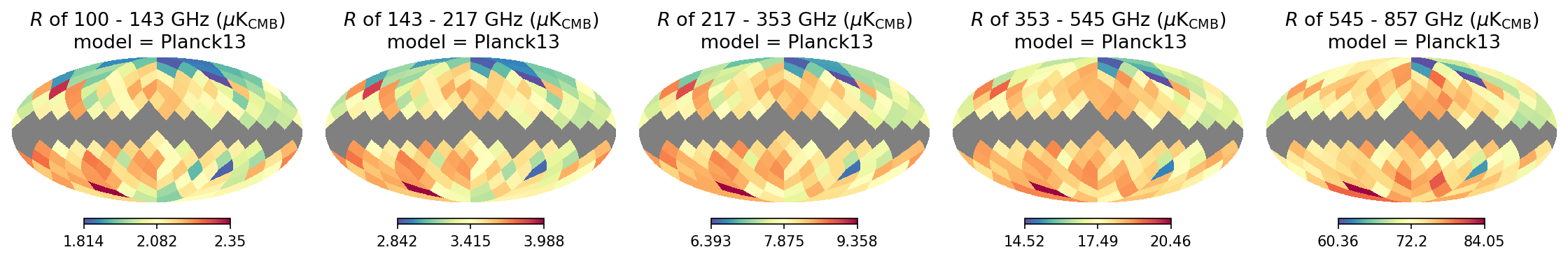}
\includegraphics[width=0.75\textwidth]{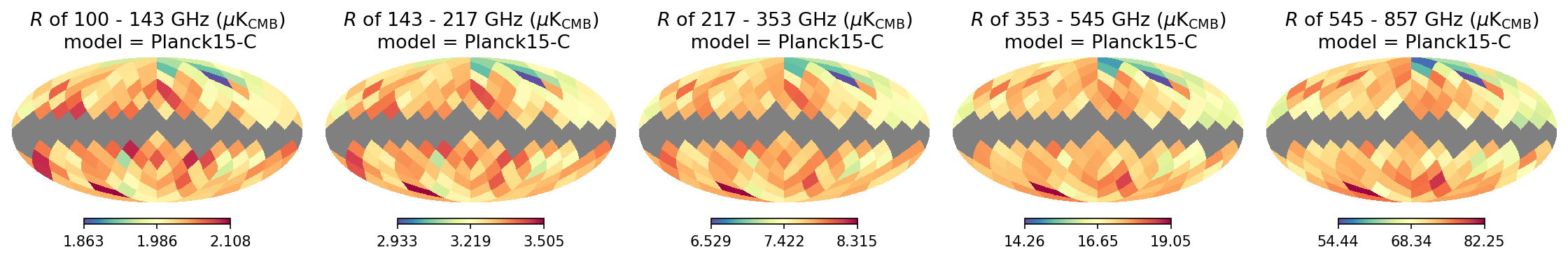}
\includegraphics[width=0.75\textwidth]{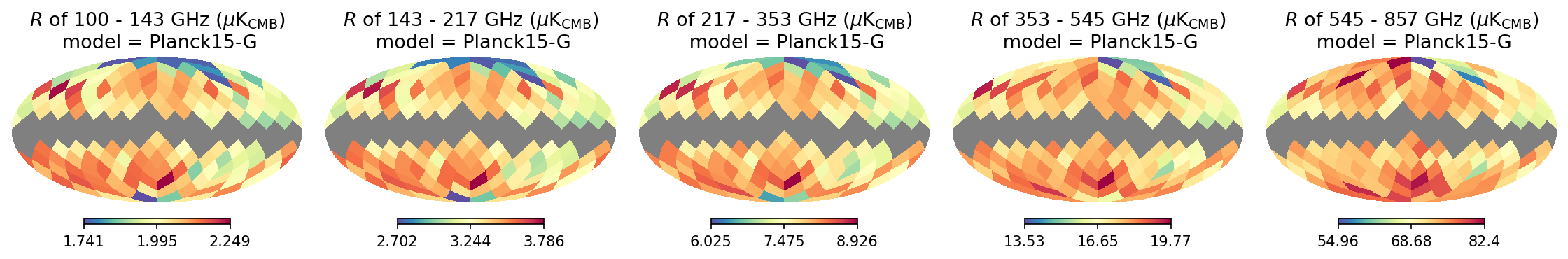}
\includegraphics[width=0.75\textwidth]{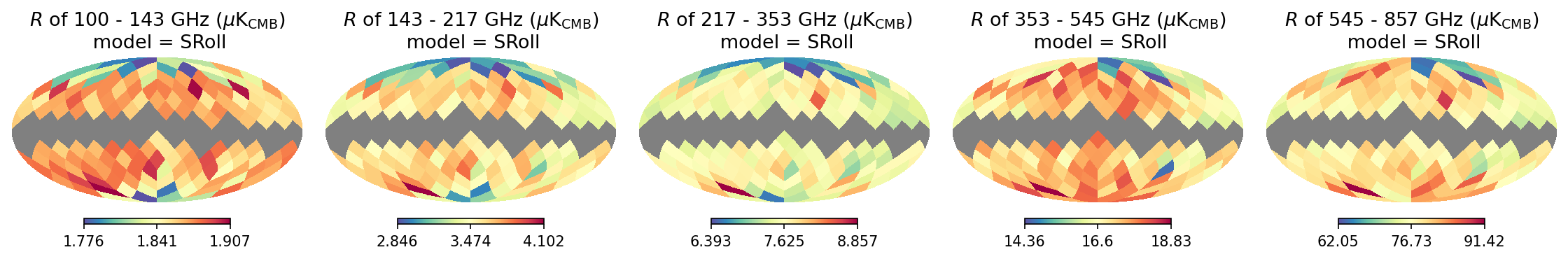}
\includegraphics[width=0.75\textwidth]{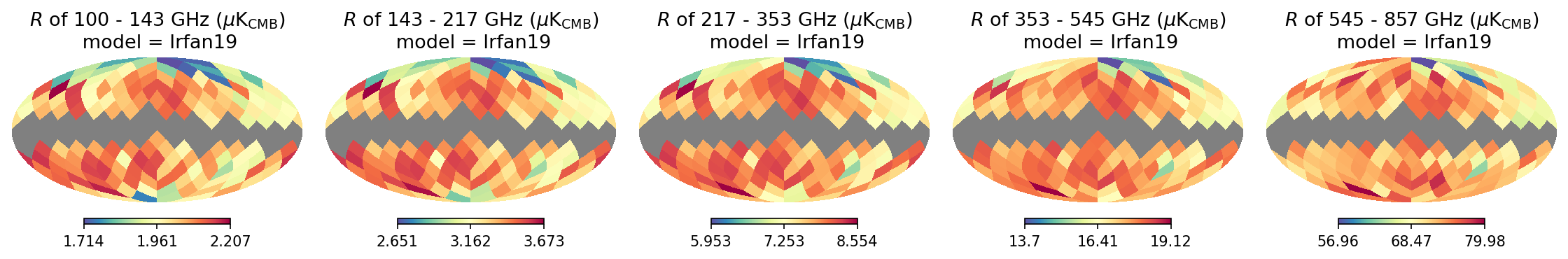}
\includegraphics[width=0.75\textwidth]{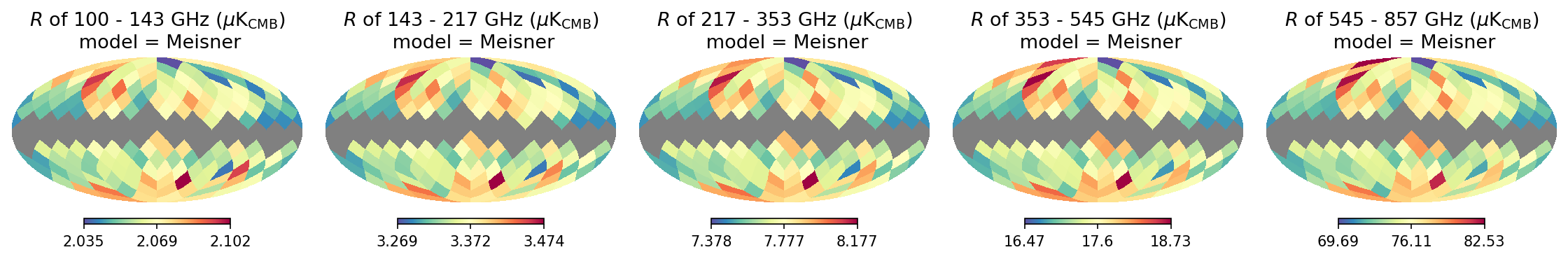}
\includegraphics[width=0.75\textwidth]{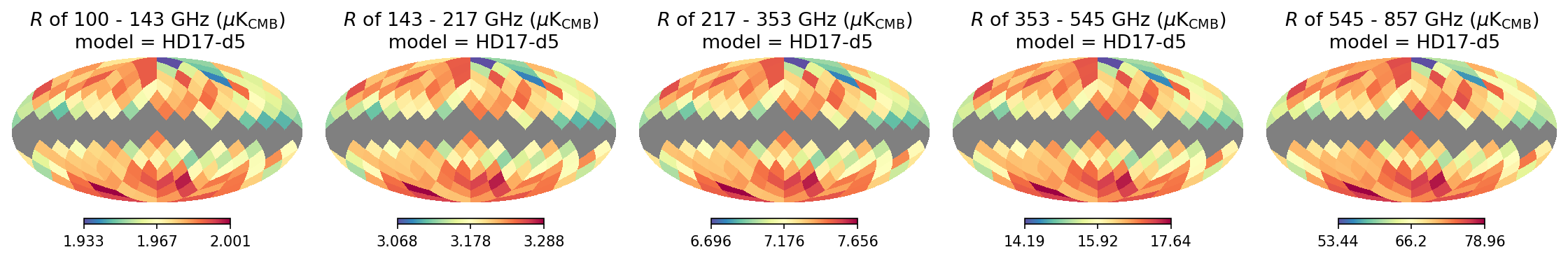}
\includegraphics[width=0.75\textwidth]{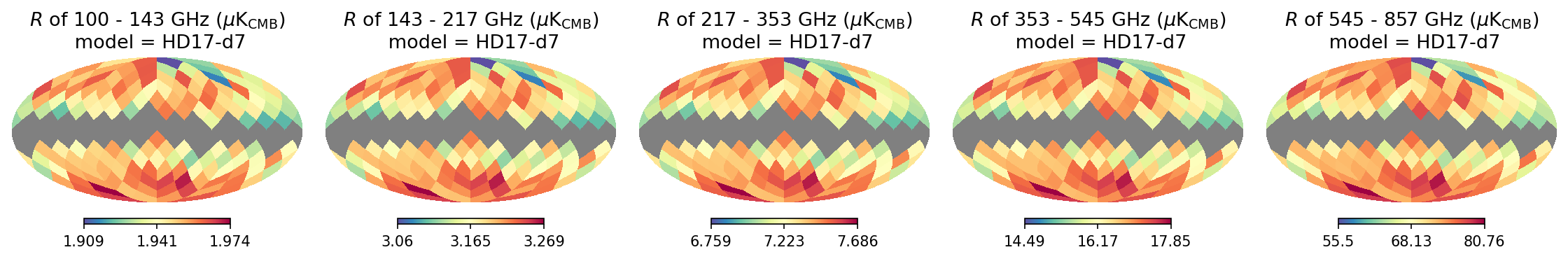}
\includegraphics[width=0.75\textwidth]{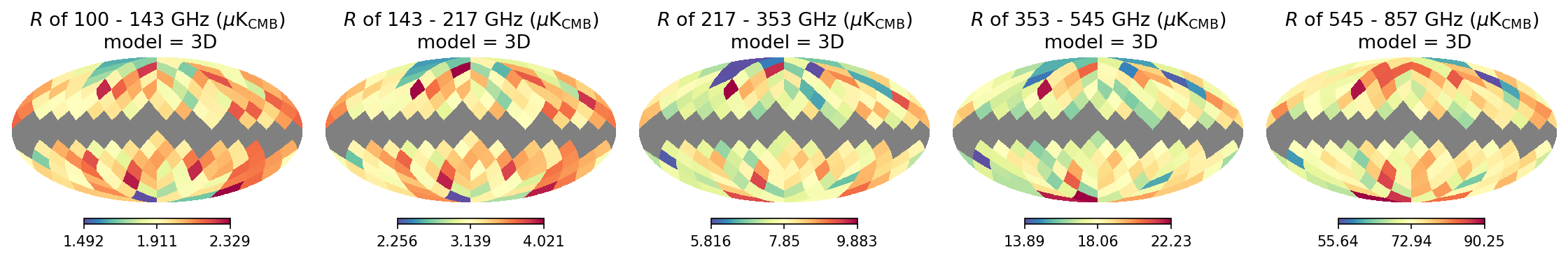}
\caption{
Maps of thermal dust intensity ratios between neighboring bands based on \textit{Planck} observed data ($R^\prime$, free from models, row 1) and based on thermal dust emission models ($R$, rows 2---10). 
From left to right: the neighboring bands for 100---143, 143---217, 217---353, 353---545, 545---857 GHz.
Horizontal comparison: For the observational data, the morphologies of intensity ratios exhibit significant variations across the 100---857 GHz range. 
However, \emph{all} models fail to provide sufficient variability to align with the observational data. 
Therefore, issues from a simple visual inspection already raise significant concerns regarding the reliability of dust emission modeling.
}
\label{fig:maps_R_thermal dust}
\end{figure}

\begin{figure}[htpb]
\centering
\includegraphics[width=0.6\textwidth]{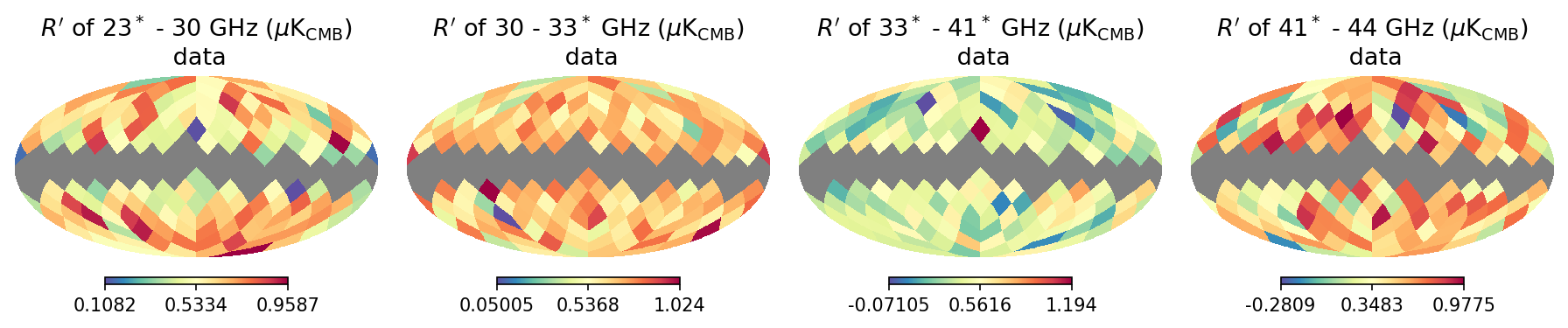}
\includegraphics[width=0.6\textwidth]{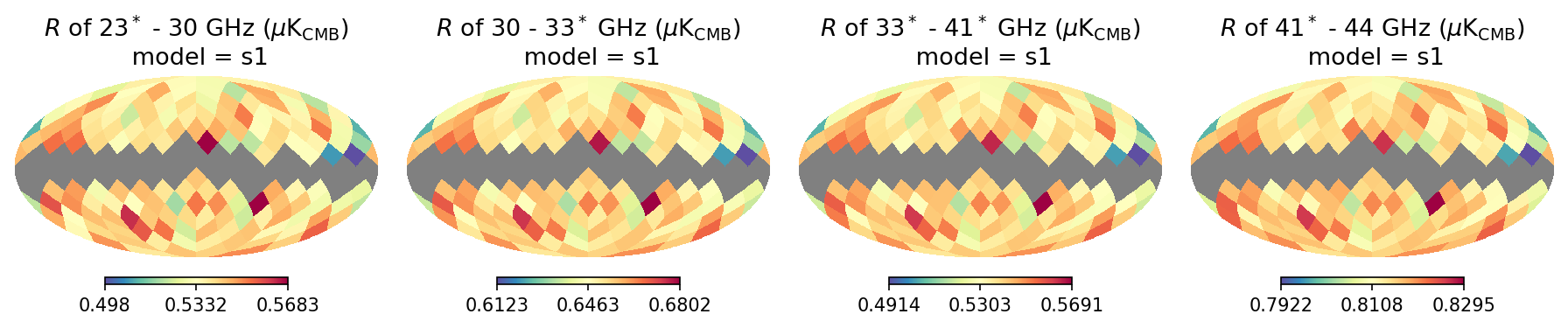}
\includegraphics[width=0.6\textwidth]{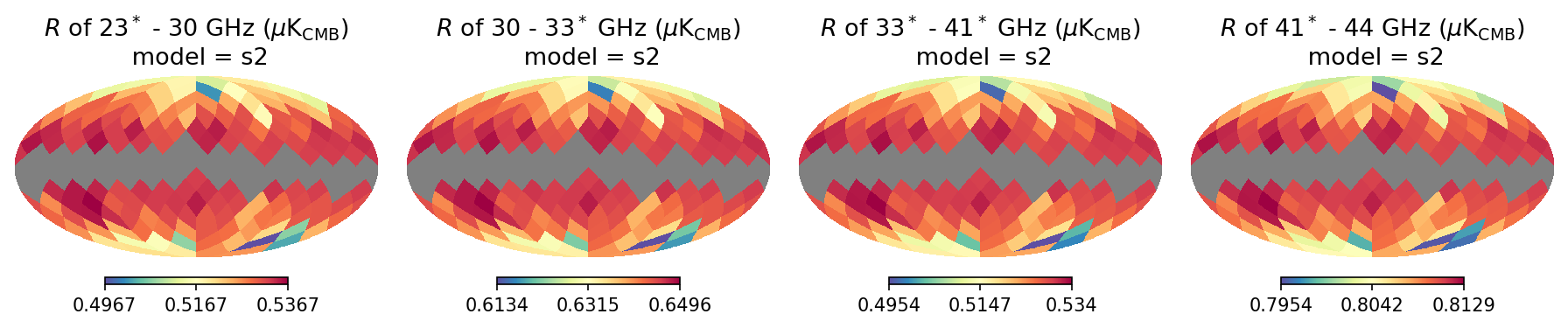}
\includegraphics[width=0.6\textwidth]{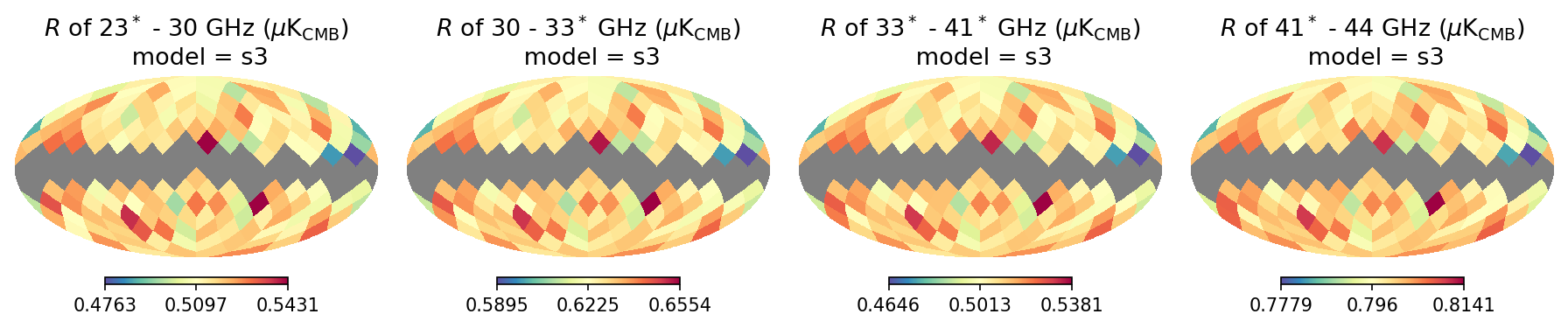}
\includegraphics[width=0.6\textwidth]{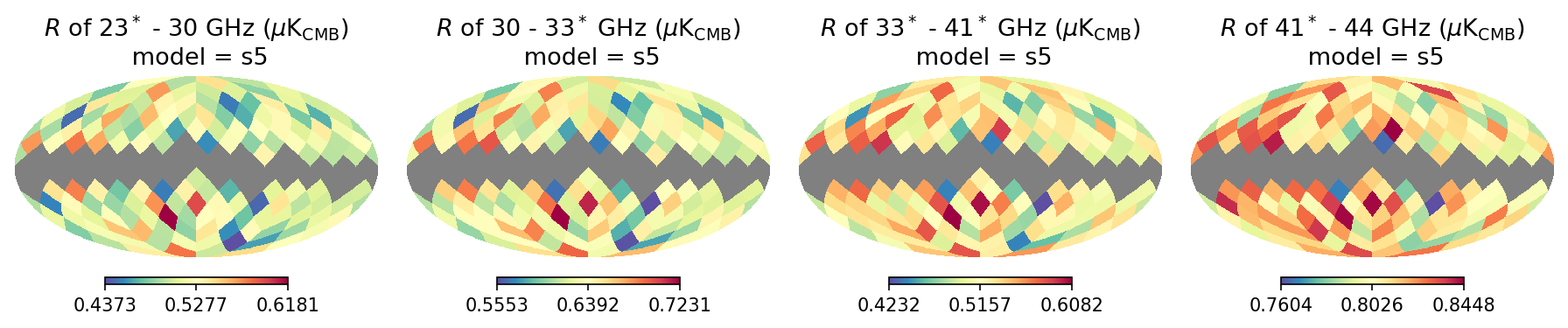}
\includegraphics[width=0.6\textwidth]{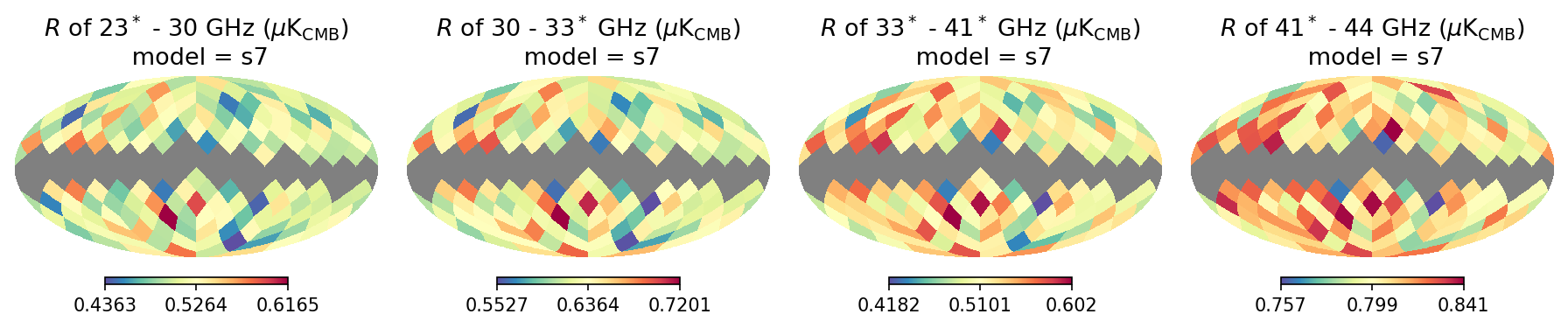}
\caption{Maps of $R^\prime$ (data, row 1) and $R$ (models, rows 2---6) for synchrotron. 
Frequency bands with star represent WMAP bands. }
\label{fig:maps_R_synchrotron}
\end{figure}

\end{document}